\documentclass{amsart}

\usepackage{amsmath, amssymb}

\usepackage[colorlinks]{hyperref} 

\newcommand{\R}{\mathbb R} 
\newcommand{\C}{\mathbb C}
\newcommand{\A}{\mathcal A}
\newcommand{\cc}[1]{\overline{#1}}

\usepackage{hyphenat}

\newcommand{\E}{\mathcal E}

\makeatletter
\newcommand{\etale}{\'etal\@ifstar{\'e}{e\xspace}}
\makeatother
\newtheorem{theorem}{Theorem}[section]
\newtheorem{lemma}[theorem]{Lemma}
\newtheorem{prop}{Proposition}
\newtheorem{conjecture} {Conjecture}
\theoremstyle{definition}
\newtheorem{definition}[theorem]{Definition}
\newtheorem{example}[theorem]{Example}

\theoremstyle{remark}
\newtheorem{remark}[theorem]{Remark}
\numberwithin{equation}{section}

\newcommand{\im}{\operatorname{Im}}

\newcommand{\id}{\operatorname{id}}

\newcommand{\wn}{\widetilde{\nabla}}
\newcommand{\wnn}{\widetilde{\nabla\!\!\!\!\nabla}}
\renewcommand{\o}{\otimes}
\renewcommand{\d}{\cdot}

\title{A Generalization of King's Equation via Noncommutative Geometry} 
\author{Gourab Bhattacharya and Maxim Kontsevich}

\address{Institut des Hautes \'{E}tudes Scientifiques, 35 route de Chartres, Laboratoire Alexander Grothendieck, F-91440, Bures-sur-Yvette, France.}
\email{bhattacharya@ihes.fr, gourabmath@gmail.com}
\email{maxim@ihes.fr}

\begin{document}


\begin{abstract}
We introduce a framework in noncommutative geometry consisting of a
$*$-algebra, a bimodule endowed with a derivation (``1-forms") and a Hermitian
structure (a ``noncommutative K\"ahler form"), and a cyclic 1-cochain whose coboundary is determined by the
previous structures. This data leads to moment map equations on the space of
connections on arbitrary finitely-generated projective Hermitian module. As particular
cases, we obtain a large class of equations in algebra (King's equations for
representations of quivers, including ADHM equations), in classical gauge
theory (Hermitian Yang-Mills equations, Hitchin equations, Bogomolny and Nahm
equations, etc.), as well as in noncommutative gauge theory by Connes, Douglas
and Schwarz.
  We also discuss Nekrasov's beautiful proposal for re-interpreting
noncommutative instantons on $\mathbb{C}^n\simeq \mathbb{R}^{2n}$ as an
infinite-dimensional solution of King's equation $$\sum_{i=1}^n [T_i^\dagger,
T_i]=\hbar\cdot n\cdot\id_{\mathcal H}$$ where $\mathcal H$ is a
Hilbert space completion of a finitely-generated $\mathbb
C[T_1,\dots,T_n]$-module (e.g. an ideal of finite codimension).
 \end{abstract}
 
 \maketitle


\section{Introduction}
There is a  remarkable similarity between self-dual Yang-Mills equations and equations introduced by King in [\ref{king}] for representations of quivers. The underlying reason is that both equations are obtained from appropriate moment maps.
 We introduce in this paper a common generalization based on  noncommutative geometry. In this setup the moment map equation is governed by a cyclic 1-cochain. Examples of a generalized King's equation include ADHM equations, noncommutative instantons, vortex equations (in particular Hitchin and Vafa-Witten equations), as well as Bogomolny and Nahm equations for the gauge group $U(k)$. Furthermore,  we discuss Nekrasov's suggestion to reinterpret noncommutative instantons as  infinite-dimensional versions of King's equation, also related to Quantum minimal surfaces considered recently in [\ref{hoppe}].

\section{Some motivations and backgrounds} 
\subsection{Mumford stability and harmonic representatives: examples}  One of major recurrent themes in K\"ahler geometry is an equivalence between the algebro-geometric property of a polystability, and the existence of a kind of harmonic metric. Let us start with several motivating examples.

\subsubsection{ Kempf-Ness Theorem.}  Let $G$ be an algebraic reductive group over $\mathbb{C}$  acting linearly on a finite dimensional vector space $V$ over  $\mathbb{C}$.

\begin{definition}
A non-zero orbit $G \cdot v \subset V - \{0\}$ is called \emph{semistable} iff its closure does not contain $0$.
\end{definition}

It is easy to see that the union of all semistable orbits forms an open $G$-invariant subset of $V$ (possibly empty).

\begin{definition}
A semistable orbit is called \emph{polystable} iff it is closed (equivalently, closed in the semistable locus).
\end{definition}

Let us choose a maximal compact subgroup $K \subset G$ and a Hermitian norm  $\lVert \cdot \rVert $ on $V$ invariant under the $K$-action. 
By definition,  on a semistable  orbit $G \cdot v $ the function $\log (\text{norm}) $ is bounded below. 

\begin{theorem} (Kempf-Ness [\ref{kempf}])
 A semistable orbit $G \cdot v$ is polystable iff the restriction of the function $\log (\operatorname{norm}) $ to this orbit achieves a minimum.  Moreover, in this case the locus of minima is a unique orbit of $K$.
\end{theorem}

The set of polystable orbits coincides with the set of $\mathbb{C}$-points of the reduced scheme $\mathcal{M} := \text{Spec}(A) - \{0\}$, where $A = \mathbb{C}[V]^G$ is the algebra of invariants. 

The function,
\begin{equation}
    H : G \cdot v \mapsto \text{min}_{g \in G} \log ( \lVert g \cdot v \rVert ) \in \mathbb{R}
\end{equation}
is a \textit{plurisubharmonic} continuous function on $\mathcal{M}$.  Moreover, on the smooth locus of $\mathcal{M}$, the function $H$ is the potential of a K\"{a}hler metric $\omega_{M} = i \partial \overline{\partial} H$.

\begin{example} Fix integers $r,n\ge 1$. 
If $G=GL(r,\C)$ (with the maximal compact subgroup $K=U(r)$) and the  representation $V$ is the direct sum of $n$ copies of the adjoint representation of $G$,  then the local minima of the function $\log (\text{norm}) $ on non-zero orbits are non-zero collections $(T_1,\dots,T_n)$ of $n$ operators in $\C^r$ satisfying
 \begin{align}\label{Kingsimple}
     \sum_{i=1}^n [T_i^\dagger, T_i]=0
 \end{align}
 where $T^\dagger_i$  is the Hermitian conjugate to $T_i$. The polystable orbits, together with the zero orbit, are exactly the conjugacy classes
  of  $r$-dimensional semisimple representations of the free algebra $\C\langle  T_1,\dots T_n\rangle$.
\end{example}

\ 

\subsubsection{King's Theorem}  A quiver is a finite oriented graph.  Here is the formal definition:  
\begin{definition}
A \emph{quiver} $Q = ( Q_0 , Q_1 , s, t ) $  is a tuple consisting of finite sets $Q_0$, $Q_1$  (whose elements are called vertices and arrows of $Q$ respectively), and two maps $s : Q_0 \rightarrow Q_1$, $t : Q_1 \rightarrow Q_0 $, (called the source and the target maps).
\end{definition}
\begin{definition}
A \emph{representation} $\mathcal{E}$ of a quiver $Q$ over a field $k$ is given by a collection of $k$-vector spaces $\mathcal{E}_v$  for each vertex $v \in Q_0$,  and a collection of morphisms $T_a : \mathcal{E}_{s(a)} \rightarrow \mathcal{E}_{t(a)} $  for each arrow $a \in Q_1$.
\end{definition}

The representations of a given quiver form an  abelian category.  
\begin{definition}
Let us fix a collection of numbers $\eta = ( \eta_v \in \mathbb{R})_{v \in Q_0 } $  associated with the vertices of $Q$.  Let $\mathcal{E}$  be a non-zero finite dimensional representation  of a quiver $Q$  such that, 
\begin{equation} \label{etaparameters}
    \sum_{v \in Q_0}   \eta_v \cdot \dim \mathcal{E}_v = 0 \in \mathbb{R}\,.
\end{equation}
Then, $\mathcal{E}$ is called \emph{semistable with slope} $\eta$ (or, equivalently $\eta$-semistable ) iff for any subrepresentation $\mathcal{E}^\prime \subset \mathcal{E}$ such that $\mathcal{E}^\prime \neq 0, \mathcal{E}$, one has $\sum_{v \in Q_0} \eta_v \cdot \dim \mathcal{E}^\prime_v \leq 0 $.  A $\eta$-semistable representation is called \emph{$\eta$-stable} iff in the previous condition one has strict inequality   $\sum_{v \in Q_0} \eta_v \cdot \dim \mathcal{E}^\prime_v < 0 $.  A $\eta$-semistable representation is called \emph{polystable} iff it is a direct sum of $\eta$-stable ones.
\end{definition}

\

For any given $\eta$, the semistable representations with slope $\eta$, together with the zero representation, form an  artinian abelian category.  The simple objects in this category are exactly the $\eta$-stable representations, whereas the non-zero semisimple objects are exactly the $\eta$-polystable representations.

\begin{theorem} (A.~D.~King [\ref{king}])
 In the case $k = \mathbb{C}$, a representation is $\eta$-polystable iff there exists a collection of Hermitian norms $( \lVert \cdot \rVert_v )_{v \in Q_0}$ on vector spaces $(\mathcal{E}_v)_{v \in Q_0}$  such that on the orthogonal direct sum $E:=\boxplus_v \mathcal{E}_v$  one has the following equality:
 \begin{equation}\label{King}
     \sum_{a \in Q_1} [T^\dagger_a, T_a] = \sum_{v \in Q_0} \eta_v \cdot \mathit{Pr}_{\mathcal{E}_v}
 \end{equation}
 taking place in the algebra of operators in $E$, where  $\text{Pr}_{\mathcal{E}_v}$  is the orthogonal projection to the direct summand $\mathcal{E}_v$.
\end{theorem}

Notice that  \eqref{King} is equivalent to a collection of individual constraints for each vertex $v\in Q_0$:
\begin{align}\label{Kingatvertex}
\forall v\in Q_0:\qquad  \sum_{a \in Q_1} \mathit{Pr}_{\mathcal{E}_v} \cdot [T^\dagger_a, T_a]\cdot \mathit{Pr}_{\mathcal{E}_v}  = \eta_v \cdot \mathit{Pr}_{\mathcal{E}_v}\in  \mathit{Pr}_{\mathcal{E}_v}  \cdot  End(E)\cdot  \mathit{Pr}_{\mathcal{E}_v}  \simeq End(\mathcal E_v) \,.
\end{align}

Similarly to the Kempf-Ness theorem, the set of isomorphism classes of $\eta$-polystable representation of $Q$ with a given dimension vector,
\begin{equation}
    \overrightarrow{\dim} (\mathcal{E}) : =  (\dim (\mathcal{E}_v)_{v \in Q_0} ) \in \mathbb{Z}^{Q_0}_{\geq 0}\,,
\end{equation}
is the set of $\mathbb{C}$-points of a reduced separated scheme over $\mathbb{C}$.  Moreover, its open dense subset of smooth points is endowed with a natural K\"{a}hler metric.  

\ 

\subsubsection{Donaldson-Uhlenbeck-Yau (DUY) Theorem}
Let $X/\mathbb{C} $ be a smooth connected K\"{a}hler manifold of complex dimension $n > 0$, and $\nu \in H^2(X;\mathbb{R}) \cap H^{1,1} (X)$ be a K\"{a}hler class.  We assume that
\begin{equation}
    \langle [X], \nu^n \rangle = 1\,.
\end{equation}
\begin{definition}
For $\lambda \in \mathbb{R}$, a holomorphic vector bundle $\mathcal{E}$ on $X$ is called \emph{$\lambda$-stable} if 
\begin{align}\langle [X], c_1(\mathcal{E}) \cdot \nu^{n-1} \rangle = \lambda \cdot \text{rank}(\mathcal{E} )\end{align} 
and for any torsion-free coherent subsheaf $0\ne \mathcal{E}^\prime \subset \mathcal{E} $ such that $\text{rank}(\mathcal{E}^\prime) < \text{rank} (\mathcal{E}) $  one has \begin{align}\label{DUYineq} \langle[X], c_1 (\mathcal{E}) \cdot \nu^{n-1} \rangle < \lambda \cdot \text{rank} (\mathcal{E}^\prime)\,. \end{align} 
 Equivalently, in \eqref{DUYineq} one can  replace torsion-free subsheaves by \emph{subbundles} of $\mathcal E$ restricted to the complements $X-Z$ to  closed analytic subsets $Z\subset X$ of complex codimension at least $2$.
A \emph{$\lambda$-polystable} bundle is defined as a finite sum of $\lambda$-stable ones.
\end{definition}

\ 

\begin{theorem}([\ref{donaldson2}], [\ref{yau}])
 For a choice of a K\"{a}hler $(1,1)$-form $\omega^{1,1}$ on $X$ with $[\omega^{1,1}] = \nu $, we have the following: a vector bundle $\mathcal{E} $ is $\lambda$-polystable iff it admits a Hermitian metric $h_{\mathcal{E}}$ such that the curvature form  $F = F_{h_{\mathcal{E}}} $ of the canonical connection associated with $h_{\mathcal{E}}$ satisfies the Hermitian Yang-Mills equation (HYM in short):
\begin{equation}\label{HYMequation}
    \frac{1}{2\pi \sqrt{-1}} F \cdot (\omega^{1,1})^{n-1} = \lambda \cdot \id_{\mathcal{E}} \cdot (\omega^{1,1})^n \in \Gamma ( X , \mathcal{E}^{\star} \otimes \mathcal{E} \otimes \Omega^{n,n}_{X} )\,.
\end{equation} 
\end{theorem}
The DUY theorem is a famous example of Kobayashi-Hitchin type correspondences in differential geometry.

Later this result was generalized in [\ref{bando}] by S.~Bando and Y.-T.~Siu to so-called \emph{reflexive} sheaves
\begin{align}\label{reflexive} \mathcal E\in Coh(X), \quad \mathcal E=\mathcal E^{**}\qquad\text {where } \mathcal E^*:=\mathcal{H}om(\mathcal E,\mathcal O_X)\,,\end{align}
which can be alternatively viewed as vector bundles defined outside of closed analytic subsets of complex  codimension at least 2.

\ 

\subsection{Geometry of moment maps}  Let $(M,\omega_M)$ be a symplectic manifold.  Let a  \emph{connected}\footnote{This is a simplifying assumption which holds in the context of our paper.} compact Lie group $K$ with Lie algebra $\mathfrak{k}$ acts smoothly on $M$ and preserves the symplectic form $\omega_M$. 
 Then we get a homomorphism 
of Lie algebras
\begin{align}u\in \mathfrak{k}\mapsto X_u\in \Gamma(M,T_M), \quad X_{[u_1,u_2]}=[X_{u_1},X_{u_2}], \quad \mathcal L_{X_u}\omega_M=0
\,.\end{align}
The condition $ \mathcal L_{X_u}\omega_M=0$ implies that the 1-form $i_{X_u}\omega_M$ is closed, as follows from the Cartan formula $\mathcal L_{X_u} =d\circ i_{X_u}+i_{X_u}\circ d$ and the closedness of $\omega_M$.

The symplectic action as above is called \emph{Hamiltonian} if a homomorphism is chosen 
\begin{align}\mathfrak K\to (C^\infty(M),\{\cdot,\cdot\}), \quad u\in \mathfrak K \mapsto H_u \end{align}
to the Lie algebra of functions on $M$ endowed with the standard Poisson bracket $\{\cdot,\cdot\}$, lifting the homomorphism  $u\mapsto X_u$ from $\mathfrak k$ to the Lie algebra of symplectic vector fields on $M$. Explicitly, it means  that
\begin{align}dH_u &=i_{X_u}\omega_M\quad  \forall u\in \mathfrak k\,,\label{momentmap1}\\
H_{[u_1,u_2]} &=\{H_{u_1},H_{u_2}\}:=\omega_M(i_{X_{u_1}},i_{X_{u_2}})\quad \forall u_1,u_2\in \mathfrak k \,.\label{momentmap2}\end{align}

The collection of Hamiltonians $(H_u)_{u\in \mathfrak k}$ gives a moment map
\begin{align} \mu:M\to{ \mathfrak k}^*,\quad x\mapsto (u\mapsto H_u(x)\in \R)\,.
\end{align} 

This is a $K$-equivariant map. We define the symplectic quotient  of $(M,\omega_M)$ for a given Hamiltonian action to be the quotient of the space  $\mu^{-1}(0)\subset M$  by the action of $K$. This quotient is a locally compact singular space in general, but it is symplectic on its open dense subset of smooth points. Moreover, if $M$ is endowed with a complex structure such that $\omega_M $ is the imaginary part of a K\"{a}hler (1,1)-form and $K$ acts by K\"{a}hler isometries, then the quotient space $\mu^{-1}(0)/K$
 is a reduced complex-analytic space with a K\"{a}hler metric on its smooth locus.
 
The constraint  $\mu(x)=0$ on a point $x\in M$ is called the \emph{moment map equation}.

 \begin{remark} \label{h1h2} For a given symplectic $K$-action, the \emph{obstruction} to the existence of a Hamiltonian lift is a class  in $H^2(\mathfrak k,\R)$. If the obstruction vanishes, then the set of all various lifts to a Hamiltonian action is  a torsor over the group of abelian characters $Hom_{Lie}(\mathfrak k, \R)=H^1(\mathfrak k,\R)$. 
 \end{remark}

\begin{example} (King's equations  as moment map equations)

Let $Q$ be a finite quiver. Fix a finite-dimensional Hermitian vector space $\E_v$ for each vertex $v\in Q_0$. Then the compact Lie group 
\begin{align} K:=\prod_{v \in Q_0 } U(\E_v)\end{align}
acts on the finite-dimensional complex vector space 
\begin{align} M:=\prod_{a\in Q_1} \text{Hom}(\E_{s(a)},\E_{t(a)})\end{align}
parameterizing representations of $Q$ in $(\E_v)_{v\in Q_0}$. We endow $M$ with the constant (i.e., translationally invariant) K\"{a}hler metric associated with the Hermitian norm on $M$ given by
\begin{align} \lVert(T_a)_{a\in Q_1}\rVert^2:=\sum_{a\in Q_1} \text{Trace}(T_a^\dagger T_a)\,.
\end{align}
The moment map in this example is given (in terms of Hamiltonians)  by the formula, where $u=(u_v)_{v\in Q_0}\in \mathfrak k$,
\begin{align}
H_u((T_a)_{a\in Q_1}):=\sqrt{-1}\cdot \text{Trace}\left(\sum_{v\in Q_0} u_v \cdot\left(\sum_{a\in Q_1} [T_a^\dagger, T_a]-\sum_{v\in Q_0} \eta_v\cdot Pr_{\E_v}\right)  \right)\,.
\end{align}
We see that the vanishing of the moment map is equivalent to King's equation \eqref{King}.

\end{example}
\begin{example} (Hermitian Yang-Mills equations as  moment map equations) 

Let $\E \rightarrow X$ be a complex vector bundle over a K\"{a}hler manifold $(X,\omega^{1,1}_X)$, endowed  with a Hermitian metric.  
We define the ``compact" group $K$ to be the group of unitary automorphisms of $\E$. The infinite-dimensional manifold $\mathcal{M}$ on which $K$ acts will be the affine space of $\cc{\partial}$-connections $\nabla^{0,1}$ on $\E$ (not necessarily integrable). The space of $\overline{\partial}$-connections has the tangent space (at each point)
 equal to $\Gamma(X,End\,\E\otimes \Omega^{0,1}_X)$, and it is endowed with the Hermitian structure given by 
\begin{align}\label{tangent}(\alpha,\beta):=\sqrt{-1}\int_X \text{Trace}(\alpha\wedge \cc{\beta})\wedge (\omega^{1,1}_X)^{\dim_\C X-1}\,.\end{align}
We define a constant (i.e. translationally invariant)  K\"{a}hler metric ${\omega}^{1,1}_{\mathcal{M}}$ on the affine space of connections by the form \eqref{tangent}
 on each tangent space. The action of group $K$ is by K\"{a}hler isometries, hence symplectic. Moreover, this action has a canonical  Hamiltonian lift, with 
 the moment map  given by 
\begin{align}
H_u(\nabla^{0,1}):=\int_X 
\text{Trace}(u\cdot \frac{1}{2\pi\sqrt{-1}}F_{\nabla^{0,1}}\cdot (\omega^{1,1}_X)^{\dim_\C X-1}
- \lambda u \cdot  (\omega^{1,1}_X)^{\dim_\C X})\,.
\end{align}

Again, we see that the vanishing of the moment map is equivalent to the HYM equation.
\end{example}

 \ 
 
\subsection{Further examples of harmonic representatives}
\subsubsection{ADHM construction}
 In physics (gauge theory) one is interested in solutions of HYM equations \eqref{HYMequation}  in the case of a \emph{non-compact} space $X=\R^4=\C^2$ endowed with the standard flat metric. The solution with finite energy $\int \lVert F \rVert^2 <\infty $ are called \emph{instantons}. A classical result [\ref{adhm}] identifies instantons for the gauge group $U(k)$ and total charge $N\in \mathbb Z_{\ge 0}$ (the second Chern class $c_2$), 
 with a conjugacy classes (under the natural action of $U(k)\times U(N)$) of solutions of the system of ADHM equations
 \begin{equation}\label{ADHMeqs}
     [\alpha, \beta] + ba = 0,
    \qquad [\alpha^\dagger,\alpha] + [\beta^\dagger,\beta] + b^\dagger b - a a^\dagger  = 0
    \end{equation}
    where 
    \begin{align}
        \alpha,\beta \in End(\C^N),\quad a \in Hom(\C^k,\C^N),\,b \in Hom(\C^N,\C^k)
    \end{align}
    satisfying the following \emph{non-degeneracy} condition:
    \begin{align}\label{nondeg}
        {\text{the stabilizer of } (a,b,\alpha,\beta)  \text{ in }  \text{ is trivial}.}
    \end{align}

 \ 

    \emph{Framed} instantons are defined as  solutions of ADHM equations satisfying the nondegeneracy condition \eqref{nondeg},  modulo the (free) action of the 
     group $U(N)$ only.
    In terms of algebraic geometry, framed instantons on $\R^4$ correspond to  polystable holomorphic vector bundles $\mathcal{E}$ on $\C P^2\supset \C^2\simeq \R^4$ with the Chern classes 
    \begin{align}
        \text{rank } \mathcal E=k, \,\,c_1(\mathcal E)=0,\,\,\langle [\C P^2],c_2(\mathcal E)\rangle =N
    \end{align}
    and with the \emph{trivialization} of the restriction of $\mathcal E$ to the projective line at infinity
     $\C P^1_\infty:=\C P^2-\C^2$. The residual action of $U(k)\subset GL(k,\C)$ is via changing the trivialization isomorphism  \begin{align}\E_{\C P^1_\infty}\simeq \C^k \otimes \mathcal O_{\C P^1_\infty}\,.\end{align}
     
     One can view instantons on $\R^4=\C^2$ as solutions of $HYM$ on $\C P^2$ for a singular K\"{a}hler metric (which is the flat metric on $\C^2$), with singularities at $\C P^1_\infty\subset \C P^2$.

 \

     (Framed) ADHM equations can be re-interpreted as King's equation for the following quiver $Q^{(k)}$.
      The set of vertices  is two-element set $\{\bf{1},\bf{2}\}$. Quiver $Q^{(k)}$ has two arrows $\alpha,\beta$ connecting vertex $\bf 1$ with itself, $k$ arrows $a_1,\dots,a_k$ connecting $\bf 2$ with $\bf 1$, and $k$ arrows $b_1,\dots,b_k$ connecting $\bf 1$ with $\bf 2$. 
      
      \ 
      
      \unitlength=1.5pt
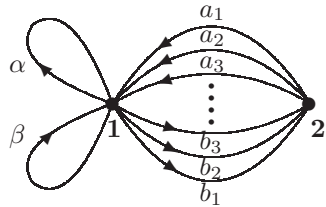
      \begin{picture}(60,60)(-120,-30)
\qbezier(0,0)(-10,27)(-20,20)
\qbezier(0,0)(-27,10)(-20,20)
\qbezier(0,0)(-10,-27)(-20,-20)
\qbezier(0,0)(-27,-10)(-20,-20)
\qbezier(0,0)(25,39)(50,0)
\qbezier(0,0)(25,27)(50,0)
\qbezier(0,0)(25,15)(50,0)
\put(22,22){$ a_1$}
\put(22,15.5){$ a_2$}
\put(22,9){$a_3$}
\qbezier(0,0)(25,-39)(50,0)
\qbezier(0,0)(25,-27)(50,0)
\qbezier(0,0)(25,-15)(50,0)
\put(22,-24.5){$ b_1$}
\put(22,-18){$b_2$}
\put(22,-11.5){$b_3$}

\put(0,0){\circle*{3}}
\put(49.5,0){\circle*{3}}

\put(25,4.5){\circle*{1.1}}
\put(25,1.5){\circle*{1.1}}
\put(25,-1.5){\circle*{1.1}}
\put(25,-4.5){\circle*{1.1}}

\thicklines
\put(-16,7.7){\vector(-1,1){3}}
\put(-17.5,-9.6){\vector(1,1){3}}
\put(14,15.5){\vector(-3,-2){3}}
\put(14,11){\vector(-2,-1){3}}
\put(14,6){\vector(-3,-1){3}}
\put(14,-15.5){\vector(3,-2){3}}
\put(14,-11){\vector(2,-1){3}}
\put(14,-6){\vector(3,-1){3}}

\put(-26,8){$\alpha$}
\put(-26,-10){$\beta$}

\put(-1.5,-8.5){$\bf 1$}
\put(50,-8.5){$\bf 2$}

\end{picture}

       A solution of ADHM equations gives a representation $\mathcal F$ of $Q^{(k)}$ in the Hermitian spaces $\mathcal F_{\bf 1}=\C^N$, $\mathcal F_{\bf 2}=\C^1$ (endowed with the standard Hermitian norm), satisfying the constraints
        \begin{equation}\label{muhol}
     [\alpha, \beta] + \sum_{i=1}^k b_i a_i = 0,
    \end{equation}
    \begin{equation}\label{muher}
     [\alpha^\dagger,\alpha] + [\beta^\dagger,\beta] + \sum_{i=1}^k  b_i^\dagger b_i - \sum_{i=1}^k a_i a_i^\dagger  = 0\,.
    \end{equation}
       Equation \eqref{muhol} can be viewed as a \emph{relation} in the path algebra of $Q^{(k)}$,
       and equation \eqref{muher} can be viewed as King's equation at vertex $\bf 1$, cf. \eqref{Kingatvertex}. Notice that the King's equation at vertex $\bf 2$ is automatically satisfied by the following reason:
     we have an obvious trace identity
     \begin{align}
         \text{Trace}\left([\alpha^\dagger,\alpha] + [\beta^\dagger,\beta] + \sum_{i=1}^k  [b_i^\dagger, b_i] +\sum_{i=1}^k [a_i^\dagger, a_i]\right)=0\,.
     \end{align}
     Therefore, equation \eqref{muher} implies that the l.h.s. of the King's equation at the vertex $\bf 2$ has also trace 0, but it is an endomorphism of the $1$-dimensional space $\mathcal F_{\bf 2}=\C^1$, hence it is equal to 0 as an operator.
     
     \subsubsection{Instantons on noncommutative $\R^4$ and deformed ADHM construction} About $20$ years ago, motivated by ideas from string theory, following pioneering work [\ref{connes}],   N.~Nekrasov and A.~Schwarz in [\ref{schwarz}] proposed a generalization of ADHM construction and HYM equations to the case of \emph{noncommutative}  flat space $\R^4_\theta$. The latter is understood as certain completion of quantum algebra $\mathcal A_\theta$ generated by coordinates $x_1,x_2,x_3,x_4$ satisfying commutation relations
     \begin{align}
         [x_i,x_j]=\sqrt{-1}\cdot \theta_{ij}
     \end{align}
     where $\theta=(\theta_{ij})_{1\le i,j\le 4}$ is a real non-degenerate skew-symmetric $4\times 4$ matrix. A bundle over the  noncommutative space, corresponding to  $\mathcal A_\theta$, is understood as a finitely-generated projective $\mathcal A_\theta$-module.
     The space of framed instantons on  noncommutative $\R^4_\theta$ is in one-to-one correspondence with the set of  solutions of the deformed ADHM equations
     \begin{align}\label{defADHMeqs}
    [\alpha, \beta] + ba = 0,
    \qquad [\alpha^\dagger,\alpha] + [\beta^\dagger,\beta] + b^\dagger b - a a^\dagger  = \eta\cdot \id_{\C^N},\quad\eta\ne 0      
     \end{align}
\emph{without} any non-degeneracy condition like \eqref{nondeg}. The deformed  ADHM equations can be (again)  interpreted as King's equations for the same  quiver $Q^{(k)}$ but with the deformed moment map (parameters $\eta_v$ as in \eqref{etaparameters}).

Each instanton on  noncommutative space $\R^4_\theta$ gives a torsion-free module $E$ over $\C[z_1,z_2]$ where $z_1,z_2$ are two complex coordinates on $\C^2\simeq \R^4$, which is extended to a coherent sheaf on $\C P^2$ trivialized as a bundle at $\C P^1_\infty$. In contrast with the commutative case, $E$ is not necessarily locally-free (i.e. not a vector bundle globally). For example, $E$ could be an ideal of finite codimension in $\C[z_1,z_2]$, giving a large class of examples of instantons of rank $k=1$ on $\R^4_\theta$ which does not have any analog in the commutative limit $\theta\to 0$. Notice that such torsion-free coherent sheaves
 are \emph{not} reflexive (see \eqref{reflexive}, hence are are excluded in the classical (commutative)
 Kobayashi-Hitchin correspondence.

\subsubsection{Nekrasov's proposal: an infinite-dimensional King's equation}

Soon after [\ref{schwarz}] it was observed in works by K.~Furuuchi [\ref{furuuchi}] and by N.~Nekrasov [\ref{nekrasov}] that the equations for an instanton on $\R^4_\theta$ for $\theta\ne 0$ are in a sense equivalent to a structure of pre-Hilbert space on $\C[z_1,z_2]$-module $E$ satisfying certain constraint which is an infinite-dimensional generalization of King's equation, which differs drastically from ADHM  equations.
 This equivalence is \emph{not} translationally invariant, in a sense it depends on a specific coherent state for algebra $
 \mathcal A_\theta$ which is ``centered'' at point $0\in \R^4$.

Many  years ago one of us (M.K) was told by N.~Nekrasov that the correspondence between solutions of HYM equations on flat  noncommutative spaces and solutions of the infinite-dimensional King's equation should exist in \emph{any} complex dimension $n$ of the flat space $\C^n\simeq \R^{2n}$, beyond the hyperk\"{a}hler case $n=2$ where we have ADHM construction at our disposal.

In what follows we will describe informally the infinite-dimensional King's equation from Nekrasov's proposal.  In the last section of the paper \ref{HYMtoKing} we will sketch a derivation of the infinite-dimensional King's equation from  HYM equations on flat  noncommutative spaces  $\R^{2n}_\theta$ for arbitrary $n$.

\ 

Let $E=E_{\text{global}}$  be a finitely generated torsion-free $\mathbb{C}[z_1,z_2, \dots , z_{n}]$-module, corresponding to an algebraic  coherent sheaf $\mathcal{E}$ on $\C P^n$ which is a vector bundle outside of a finite set of points in $\C^n\subset \C P^n$, together  with the trivialized restriction to $\C P^{n-1}_\infty:=\C P^n-\C^n$.

 The infinite-dimensional King's-like equation (which we suggest to call \emph{Nekrasov equation}) is the equation on a positive Hermitian inner product $h = h_{\text{global}}$ on ${E}_{\text{global}}$.  Let us denote by $\mathcal{H} = \mathcal{H}_{h}$  the completion of the vector space  ${E}_{\text{global}}$ with respect to $h$.  The action of generators $z_i \in \mathbb{C}[z_1,z_2, \cdot \cdot \cdot , z_{n}] $ give rise to commuting unbounded operators $Z_i$ on $\mathcal{H}$.  The proposed equation is,
\begin{equation}\label{Nekrasov}
    \sum_{i=1}^n [Z_i^\dagger , Z_i] = \hbar \cdot n \cdot \id_{\mathcal{H}}
\end{equation}
where the "Planck's constant" $\hbar > 0$ is only a real parameter, and Hermitian conjugates $Z_i^\dagger$ are taken with respect to $h$.

We cannot help but ask the reader to notice the remarkable similarity between King's equation  \eqref{Kingsimple} (for the quiver with one vertex and $n$ loops) and Nekrasov equation \eqref{Nekrasov}.

This is not yet a precise mathematical formulation because one should specify the "behaviour at infinity".  Presumably, it is given by the condition
\begin{equation}
     \forall\, 1\le i,j \le n:\quad [Z_i^\dagger , Z_j] = \hbar \delta_{ij}  \cdot \id_{\mathcal{H}} + \text{trace class operator}\,.
\end{equation}

Also, Nekrasov argued that for torsion-free algebraic coherent sheaves on $\mathbb{C}^n$ of higher ($k>1$) rank, the solutions of noncommutative HYM should approximate the solutions of the  usual HYM equation in the limit $\hbar \rightarrow 0$, at least at the open locus in $\mathbb{C}^n$ where the sheaf is a bundle.  First, the space of positive Hermitian products on ${E}_{\text{global}}$  is an approximation to the space of Hermitian metrics on a holomorphic vector bundle over $X$.  Indeed, e.g. for $\mathcal{E} = \mathcal{O}^{\oplus k}_{\C^n}$ the Hermitian product on ${E}_{\text{global}}=\C^k\otimes  \mathbb{C}[z_1,z_2, \cdot \cdot \cdot , z_{n}]$  is given (roughly) by a positive self-adjoint element in 
\begin{equation}
 {E}_{\text{global}} \hspace{1mm} \widehat{\otimes} \hspace{1mm} \overline{{E}}_{\text{global}} = \mathbb{C}[z_1,z_2, \cdot \cdot \cdot , z_{n}] \hspace{1mm} \widehat{\otimes} \hspace{1mm} \mathbb{C}[\cc {z}_1,\cc{z}_2, \cdot \cdot \cdot , \cc{z}_{n}] \otimes(\C^k\otimes \overline{\C}^k)\cong C^\infty (\mathbb{R}^{2n})\otimes_\R Mat(k\times k,\C),
\end{equation}
and then should give a metric in the trivial bundle of rank $k$ on $\mathbb{C}^n$.

\ 

Following two (informal) conjectures are due to Nekrasov.

\begin{conjecture}
Equation \eqref{Nekrasov} has a unique solution with a given appropriate boundary condition at infinity.
\end{conjecture}

\begin{conjecture} 
In the limit $\hbar \rightarrow 0$ solutions of the equation \eqref{Nekrasov} approaches to the solutions of the equation \eqref{HYMequation} with parameter $\lambda=0$.
\end{conjecture}

It seems that one can generalize all this to \emph{arbitrary} coherent sheaves on $\C^n$, not necessarily torsion-free. Presumably, the sheaf should be pure of certain dimension $m\le n$ (meaning that the dimension of support of the sheaf is $m$, and the sheaf has no non-zero subsheaves with at most $(m-1)$-dimensional support). Moreover, the trivialization at infinity (in the case $m=n$) should be replaced by an \emph{extension} to $\C P^{n-1}_\infty$ together with a metric on it satisfying HYM equation.  The corresponding Nekrasov equation is
\begin{equation}\label{Nekrasovnew}
    \sum_{i=1}^n [Z_i^\dagger , Z_i] = \hbar \cdot m \cdot \id_{\mathcal{H}}\,.
\end{equation}
As an example we mention King's equation for finite-dimensional representations of $\C[z_1,\dots,z_n]$ (the case $m=0$, the equation is literally the same as \eqref{Kingsimple}), and the case $m=1$ for curves in affine spaces 
studied partially before (see [\ref{hoppe}] and references therein).

\ 

\section{Algebraic formalism: synopsis}

Let us fix some notations.
 For an associative unital algebra $\A$ over $\C$, we denote by $\cc{\A}$ the complex-conjugate algebra:
\begin{equation} \cc{f}+\cc{g}=\cc{f+g}
,\quad  \cc{f}\cdot \cc{g}=\cc{f\cdot g}, 
\quad \cc{\lambda f}=\cc{\lambda}\cdot \cc{f}\qquad \forall f,g\in \A, \forall \lambda \in \C\,,
\end{equation}
 and by $\A^{op}$ the opposite algebra
\begin{equation} {f}^{op}+{g}^{op}={(f+g)}^{op}
,\quad  {f}^{op}\cdot {g}^{op}={(g\cdot f)}^{op}, 
\quad {(\lambda f)}^{op}={\lambda}\cdot {f}^{op}\qquad \forall f,g\in \A, \forall \lambda \in \C\,.
\end{equation}
There are canonical isomorphisms
\begin{equation}
(\A_1\otimes \A_2)^{op}\simeq \A_1^{op}\otimes \A_2^{op}, \quad \cc{\A_1\otimes \A_2}\simeq \cc{\A_1}\otimes\cc{ \A_2},\quad \cc{\A^{op}}\simeq \cc{\A}^{op}, \quad \cc{\cc{\A}}\simeq (\A^{op})^{op}\simeq\A\,.
\end{equation}
If $E$ is a left module over $\A$ then $\cc{E}$ is a left module over $\cc{\A}$. Similarly, a \emph{right} module over $\A$ is the same as a \emph{left} module over $\A^{op}$. We have a duality between finitely-generated projective left module $E$ over $\A$ and  finitely-generated projective right modules
\begin{equation} E \leftrightsquigarrow E^\vee:=Hom_{\A-mod}(E,\A)\in mod-\A,\quad E=Hom_{mod-\A}(E^\vee, \A)\,.
\end{equation}

\

A $*$-algebra is an associative unital algebra $\A$ over $\C$ endowed with an anti-linear involution $f\mapsto f^*$ satisfying
\begin{equation} (f^*)^*=f,\,\, {f}^{*}+{g}^{*}={(f+g)}^{*}
,\,\,  {f}^{*}\cdot {g}^{*}={(g\cdot f)}^{*}, 
\,\, {(\lambda \cdot f)}^{*}={\overline{\lambda}}\cdot {f}^{*}\qquad \forall f,g\in \A, \forall \lambda \in \C\,.
\end{equation}
For any $*$-algebra $\A$ we have a canonical isomorphism $\cc{\A}\simeq {\A}^{op},\quad \cc{f}\mapsto (f^*)^{op}$. 
An element $f\in C$ is called \emph{Hermitian} if $f=f^*$, and  \emph{non-negative} iff it can be written as a finite sum of the form $\sum_i f_i f^*_i$.

\ 

In particular, for a $*$-algebra $\A$ and a bimodule $B$ over $\A$ (i.e. a module over $\A\otimes \A^{op}$, we can write $B\in \A-mod-\A$), the complex-conjugate $\cc{B}$ (which is a module over $\cc{\A\otimes \A^{op}}$) is naturally again a bimodule over $\A$ via the chain of canonical isomorphisms of algebras
\begin{equation}
\cc{\A\otimes \A^{op}}\simeq \cc{\A}\otimes \cc{\A}^{op}\simeq \A^{op}\otimes \A\simeq \A\otimes \A^{op}\,.
\end{equation}

\

The setup (in which later we will define the moment map equations) is the following:\newline
we are given
\begin{itemize} 
\item[\bf (A1)] an associative unital $*$-algebra $\A$ over $\C$,
\item[\bf (A2)] a bimodule $\Omega^1$ over $\A$,
\item[\bf (A3)] a derivation $d:\A\to \Omega^1$, i.e. a $\C$-linear map $d$ satisfying the Leibniz rule 
\begin{align} d(f \cdot g)= f\cdot d(g)+d(f)\cdot g,\quad \forall f,g\in  \A\label{Leibniz}\,\,,\end{align} 
\item [\bf (A4)] a bilinear form $\omega:\Omega^1\otimes_\C \overline{\Omega^1}\to \C$ (a ``noncommutative K\"ahler form") satisfying the properties 
\begin{equation}\label{kahlerpositive}\omega(\alpha,\overline{\beta})=\overline{\omega(\beta,\overline{\alpha})},\quad\omega (f\cdot\alpha\cdot g, \overline{\beta})=\omega(\alpha,\overline{f^*\cdot\beta\cdot g^*}),\quad \omega(\alpha,\cc{\alpha})>0\,\,\,\forall \alpha\ne 0\,,\end{equation} 
\item[\bf (A5)] a linear functional $\eta:\A\to \C$ satisfying 
\begin{equation}\label{cocycle}\eta(f^*)=-\overline{\eta(f)}, \quad \eta([f,g])=\frac{-1}{2\sqrt{-1}}\cdot\Bigl(\omega(df,\cc{d(g^*)})-\omega(dg, \cc{d(f^*)})\Bigr)\,.\end{equation}
\end{itemize}

\

This setup will be applied to
\begin{itemize} 
\item[\bf (M1)] a finitely-generated projective $\A$-module $E$,
\item[\bf(M2)] a \emph{connection} on $E$ which is defined as a $\C$-linear map $\nabla:E\to \Omega^1\otimes_\A E$ satisfying 
\begin{equation}\nabla(f\cdot  \phi)=df\otimes \phi +f\cdot \nabla(\phi),\quad \forall f\in \A, \,\phi\in E\label{connection}\,,
\end{equation}
\item[\bf(M3)]  a \emph{Hermitian form} on $E$ which is defined  to be a bilinear map $\mathsf H:E \otimes_\C \overline{E}\to \A$ satisfying
\begin{equation}\label{notionhermitian} \mathsf H(f \phi_1,\cc{g\cdot \phi_2})=f\cdot \mathsf H(\phi_1,\cc{\phi_2})\cdot g^*
\end{equation}
and such that the induced morphism of right modules over $\A$
\begin{align}\label{hermiso}
    \cc{E}\to E^\vee=Hom_{\A-mod}(E,\A),\quad \cc{\phi_2}\mapsto \left(\phi_1\mapsto H(\phi_1,\cc{\phi_2})\right)
\end{align} is an \emph{isomorphism} and is \emph{positive-definite}, in the sense $\mathsf H(\phi,\cc{\phi})\ge 0$ for all $\phi\in E$.
\end{itemize}

We will explain in the next section (see Proposition \ref{propaction}) that the action of the gauge group of unitary automorphisms of $E$ on the space of connections on $E$ can be lifted using \eqref{cocycle} to a \emph{Hamiltonian } action.
In particular, we will get the notion of a \emph{harmonic} representative.

\begin{definition} For  a finitely-generated projective $\A$-module $\E$ endowed with connection $\nabla$, a Hermitian form $\mathsf H$ is called {\bf harmonic} iff it satisfies the moment map equation \eqref{universalmmequation} defined later in section \ref{sectionuniversal}.
\end{definition}
\begin{remark}
Our setup \emph{differs} from the one proposed in [\ref{haiden}]. It would be interesting to compare two formalisms.
\end{remark}

\section{Explanations in two basic examples}

We will illustrate our axiomatics in the case of a quiver, or a compact $C^\infty$-manifold $X$.

\

{\bf (A1)}+{\bf(M1)}: The algebra $\A$ is  either a finite sum $\C^{Q_0}$ of copies of $\C$ (quiver case), or the algebra $C^\infty_\C(X):=C^\infty(X)\otimes_\R \C$ of smooth $\C$-valued functions on a manifold $X$, with the  involution $*$ given by the complex conjugation. In these examples $\A$ happen to be  commutative, although this property does not play any role in the general formalism. In the noncommutative gauge theory the algebra $\A$ is the algebra of functions on a noncommutative deformation of  $\R^4$.

In general, a finitely-generated projective $\A$-module $E$ is a left $\A$-module which is isomorphic 
to $\A^n\cdot P$ where $P\in Mat(n\times n,\A)$ is a projector, $P^2=P$.

  Such a module  is the same data as a collection of finite-dimensional complex vector spaces $(\E_v)_{v\in Q_0}$ where $E:=\oplus_v \E_v$ (quiver case), or the same data as a finite-dimensional complex vector bundle $\mathcal E$ over $X$ where $E=\Gamma(X,\mathcal E)$ (manifold case).

\

{\bf (A2)}: The bimodule $\Omega^1$ in the quiver case is the complex vector space $\C^{Q_1}$ spanned by the set of arrows $Q_1$ of the quiver, with the  structure of a bimodule over $\A$ given by 
 \begin{equation} a=\pi_{s(a)}\cdot a \cdot \pi_{t(a)}\,,
\end{equation}
where $\pi_v\in \A=\C^{Q_0}$ denotes the projector (the  base vector) corresponding to arbitrary $v\in Q_0$.

In the case of a manifold, the  bimodule $\Omega^1$ is the space of complex-valued 1-forms on $X$ with both the left and the right action given by the point-wise multiplication. More generally, one can consider pairs $(X,\mathcal F)$ where $\mathcal F\subset T_X\otimes_\R \C$ is a complex vector subbundle of the complexified tangent bundle $T_X$ to $X$ such that 
\begin{equation}\mathcal F+\cc{\mathcal F}=T_X\otimes_\R \C\label{big}\,.
\end{equation}
We define in this case the bimodule $\Omega^1$ as the space of sections  of the dual bundle $\Gamma(X,\mathcal F^*)$, which is the quotient of the space $\Gamma(X,T^*_X\otimes_\R \C)$ of complex-valued  1-forms on $X$.

The condition \eqref{big} is satisfied e.g. when $X$ is endowed with a complex structure and $\mathcal F=T^{0,1}_X$. More generally, the case when \eqref{big} is satisfied and $\mathcal F$ is formally integrable (which means that $\Gamma(X,\mathcal F)\subset  \Gamma(X,T_X\otimes_\R \C)$ is closed under the Lie bracket), corresponds to a foliation on $X$ with a transversal holomorphic structure. The foliation is given by the real distribution $\mathcal F \cap T_X$. In this case  the sheaf of functions on $X$ are killed by all the complex-valued vector fields which are local sections of $\mathcal F$, is the same as the sheaf of functions which are locally constant along the foliation and holomorphic on the complex quotient.

In what follows, we will call the case $\Omega^1=\Gamma(X,T^*_X\otimes_\R \C)$ the \emph{totally real case}, and the case $\Omega^1=\Gamma(X,(T^{0,1}_X)^*)$ when $X$ is endowed with a complex structure, the \emph{totally complex case}.

\

{\bf (A3)}+{\bf(M2)}: the derivation $d$ is equal to zero in the quiver case, and to the de Rham differential in the manifold case when $\Omega^1= \Gamma(X,T^*_X\otimes_\R \C)$. More generally, in the case of a complex distribution $\mathcal F$ as above, the differential $d$ is the composition of the de Rham differential $\A=C^\infty_\C (X)\to \Gamma(X,T^*_X\otimes_\R \C)$ and of the projection $\Gamma(X,T^*_X\otimes_\R  \C)\to \Gamma(X,\mathcal F^*)$.

In the quiver case, a connection on a finitely-generated projective module $E=(\E_v)_{v\in Q_0}$ is the same as an action of arrows 
\begin{equation} T_a: \E_{s(a)}\to \E_{t(a)}\quad \forall a\in Q_1
\end{equation}
which extend to an action of the path algebra of the quiver.

In the manifold case, a connection is the usual connection on a complex vector bundle, or a connection along distribution $\mathcal F$. In the totally complex case when $\mathcal F=T^{0,1}_X$, the connection in algebraic sense is the same as $\cc{\partial}$-connection on $\mathcal E$.

In the general algebraic setup, the differential $d:\A\to \Omega^1$ gives rise to a structure of a bimodule on $B:=\A\oplus \Omega^1$ given by 
\begin{equation}f\cdot (h,\alpha)\cdot g:=(f \cdot h \cdot g\,,\,\,f\cdot \alpha\cdot g+df \cdot h \cdot g),\quad \forall  f,h,g \in \A,\, \alpha\in \Omega^1\label{bimoduleB}
\end{equation}
endowed with an epimorphism $\pi_B$ onto the diagonal bimodule $\A_{diag}$ given by $(h,\alpha)\mapsto h$, and a splitting $h\mapsto (h,0)$ which is a monomorphism $i_B$ of  \emph{right} modules over $\A$. Conversely, any $\A$-bimodule $B$ together with morphisms\footnote{In the formulation of the notion of a connection belowe, the homomorphism $i_B$ plays no role. It can be completely omitted. What we really need is just a bimodule $B$ and a morphism $\pi_B:B\to \A_{daig}$ of bimodules.}
\begin{equation}\pi_B\in \text{Hom}_{\A-\text{mod}-\A}(B,\A_{diag}),\quad i_B\in \text{Hom}_{\text{mod}-\A}(\A_{diag},B)\label{bimoduleBmor}
\end{equation}
such that $\pi_B\circ i_B=id_\A$, is the same data as a bimodule $\Omega^1:=Ker(\pi_B)$ together with a derivation $d:\A\to \Omega^1$ satisfying the Leibniz rule \eqref{Leibniz}. The notion of a connection satisfying the analogous condition \eqref{connection}  can be rephrased as a homomorphism of left $\A$-modules 
\begin{equation} \widetilde{\nabla}: E\to B\otimes_\A E, \quad \widetilde{\nabla}\in \text{Hom}_{\A-\text{mod}}(E,B\otimes_\A E)\label{nablaprim}
\end{equation}
satisfying the constraint
\begin{equation}(\pi_B\otimes id_E)\circ  \widetilde{\nabla}: E\to \A_{diag}\otimes_\A E\simeq E\quad\text{ is equal to } \id_E\,.\label{nablaprimconstraint}
\end{equation}
Explicitly, the correspondence is given by
\begin{align}
  \text{connection }\nabla \rightsquigarrow \text{ morphism } \widetilde{\nabla}:\phi\mapsto (\phi,\nabla(\phi))\in E\oplus (\Omega^1\otimes_\A E)=B\otimes_\A E \,.
\end{align}
Assume that $B$ is a finitely-generated projective when considered as a right module over $\A$ (equivalently, one can replace $B$ by $\Omega^1$ because $\Omega^1\oplus \A_{diag}\simeq B$ in $\text{mod}-\A$). Then $B$ can be represented as the dual to a finitely-generated projective \emph{left} $\A$-module which we denote by $\mathit{Diff}_{\le 1}$:
\begin{equation}  \mathit{Diff}_{\le 1}\simeq \text{Hom}_{\text{mod}-\A}(B,\A),\quad B\simeq \text{Hom}_{\A-\text{mod}}(\mathit{Diff}_{\le 1},\A)
\end{equation}
In the manifold case and $\Omega^1=\Gamma(X,T_X^*\otimes_\R\C)$ the space  $\mathit{Diff}_{\le 1}$ can be naturally identified with the space of differential operators of order $\le 1$, hence the notation.

 The left $\A$-action on $B$ gives a right action on $\mathit{Diff}_{\le 1}$, therefore we have  $
\mathit{Diff}_{\le 1}\in \A-\text{mod}-\A$. The epimorphism $\pi_B$ gives (by duality) a monomorphism of bimodules $\pi^\vee_B: \A_{diag}\to \mathit{Diff}_{\le 1}$. We define the algebra $\mathit{Diff}$ of ``noncommutative differential operators'' as the quotient of the tensor algebra
\begin{equation}T_\A(\mathit{Diff}_{\le 1}):=\A\oplus \mathit{Diff}_{\le 1} \oplus (\mathit{Diff}_{\le 1}\otimes_\A \text{Diff}_{\le 1})\oplus\dots
\end{equation}
by the two-sided ideal generated by the subspace 
\begin{equation}
\{f-\pi_B^\vee(f)\,|\, f \in \A\}\subset \A\oplus \mathit{Diff}_{\le 1}\subset T_\A(\mathit{Diff}_{\le 1})\,.
\end{equation}
The algebra $\mathit{Diff}$ is filtered (with the component $\mathit{Diff}_{\le n}\subset \mathit{Diff}$ defined as the image of the subspace $\mathit{Diff}_{\le 1}^{\otimes_\A n}\subset
T_\A(\mathit{Diff}_{\le 1})$), and endowed with a homomorphism $\A\to \mathit{Diff}$.
It follows from definitions that finitely-generated $\A$-modules with connections can be identified with $\mathit{Diff}$-modules which are finitely-generated projective as  $\A$-modules. The algebra $\mathit{Diff}$ is the usual path algebra in the quiver case, and a ``free analog'' of the algebra of differential operators in the manifold case. In the totally real case $\Omega^1=\Gamma(X,T_X^*\otimes_\R\C)$, in the local coordinates $(x_1,\dots,x_k)$ on $X$,  an element of $\mathit{Diff}$  can be written as a finite sum
\begin{equation}
\sum_{l\le N, i_1,\dots,i_l\in \{1,\dots,k\}} f_{i_1,\dots,i_l} \cdot \partial_{i_1}\cdot \dots\cdot \partial_{i_l}\qquad \text{ for some }N<\infty, \quad f_{i_1,\dots, i_l}\in C^\infty_\C(X)\,,
\end{equation} 
where $\partial_i$ are free \emph{noncommutative} variables obeying the exchange relation with the elements of $C$:
\begin{equation}
\partial_i\cdot f -f \cdot \partial_i=\frac{\partial f}{\partial x_i}\in \A=C^\infty_\C(X)\,.
\end{equation}
In the totally complex case  one replaces free variables  $(\partial_i=\partial_{x_i})_{i=1,\dots,\dim_\R X}$ by the antiholomorphic derivatives $(\partial_{\cc{z}_i})_{i=1,\dots,\dim_\C X}$.

If we are interested e.g. in \emph{flat} connections (or bundles with a \emph{holomorphic} structure in the complex case), we should impose certain additional relations in $\mathit{Diff}$ (e.g. the commutativity relation $\partial_i\cdot \partial_j=\partial_j\cdot \partial_i$). The corresponding \emph{quotient} algebra is either the usual algebra of (complex-valued) differential operators in the totally real case, or its subalgebra of differential operators in $\cc{\partial}$-direction in the totally complex case.

\

{\bf (A4)}: In the quiver case, a choice of $\omega$ is equivalent to a choice of a  collection of Hermitian norms on vector spaces 
\begin{align} 
\Omega^1_{v_1,v_2}:=\pi_{v_1}\cdot \Omega^1\cdot \pi_{v_2}=\C^{\{a\in Q_1\,|\,s(a)=v_1,t(a)=v_2\}}
\end{align}
for all pairs $(v_1,v_2)$ of vertices of $Q$. For example, one can declare the generating set 
$${\{a\in Q_1\,|\,s(a)=v_1,t(a)=v_2\}}$$
to be an orthonormal basis of $\Omega^1_{v_1,v_2}$.

In the manifold case, the choice of  $\omega$ is equivalent to a choice of a Hermitian form on the vector bundle $\mathcal F\subset T_X\otimes_\R\C$. In the totally real (resp. totally complex) cases, a particular choice of such a form is given by a Riemannian metric (resp. a K\"{a}hler metric) on $X$.

\

{\bf (A5)}:  Let us denote by $d^\dagger$ the derivation $\A\to \cc{\Omega^1}$ given by 
\begin{align}\label{daggerderivation}
d^\dagger (f):=-\cc{d \,f^*}\,.\end{align}
Two derivations $d,d^\dagger$ with values in $\A$-bimodules $\Omega^1,\cc{\Omega^1}$ and a linear map $\omega:\Omega^1\otimes \cc{\Omega^1}\to \C$
satisfying 
\begin{align}\label{twoside}\omega((f\cdot \alpha \cdot g) \otimes\alpha')=  \omega( \alpha \otimes (g\cdot \alpha'\cdot f)),\quad \alpha\in \Omega^1,\alpha' \in \cc{\Omega^1},\,\, f,g\in \A\end{align}
give rise to a \emph{skew-symmetric} functional on $\A$
\begin{align}   \Psi(f\otimes g):= \omega(df \otimes d^\dagger g)-\omega(dg\otimes d^\dagger f)\end{align}
satisfying an additional reality constraint
\begin{align}\label{cocycle-reality}
    \cc{\Psi(f\otimes g)}=-\Psi(f^*\otimes g^*)\,.
\end{align}

\begin{lemma} The functional $\Psi$ satisfies the identity
\begin{align} 
\Psi(f_0 f_1 \otimes f_2)+\Psi(f_1 f_2 \otimes f_0)+\Psi(f_2 f_1 \otimes f_1)=0\,.
\end{align}
\end{lemma}
{\it Proof}: A direct calculation using \eqref{Leibniz} and \eqref{twoside} gives
\begin{multline}
    \Psi(f_0 f_1 \otimes f_2)+\Psi(f_1 f_2 \otimes f_0)+\Psi(f_2 f_1 \otimes f_1)=\Psi(f_0 f_1 \otimes f_2)+\dots=   \\
  =\omega(df_0 \,f_1 \otimes d^\dagger f_2)+\omega(f_0\, df_1 \otimes d^\dagger   f_2) -\omega(df_2\otimes d^\dagger  f_0\, f_1)-
  \omega(df_2\otimes f_0 \,d^\dagger  f_1)+\dots=\\
  =\omega(df_0\otimes f_1\, d^\dagger  f_2)+\omega(df_1\otimes d^\dagger  f_2\, f_0)-\omega(df_2\otimes d^\dagger  f_0\, f_1)-
  \omega(df_2\otimes f_0 \,d^\dagger  f_1)+\dots=0
  \end{multline}
where  triple dots in each line denote terms obtain by cyclic permutation of indices $0\to 1\to 2\to 0$. $\blacksquare$

\

So, we see  that $\Psi$ is a 2-cocycle in the cyclic cochain complex of $\A$. Recall that the latter is  defined by
\begin{align}C^n_{cycl}(\A):= \{\psi: \A^{\otimes n}\to \C\,|\,\psi(f_2\otimes\dots \otimes f_n\otimes f_1)=(-1)^{n-1}\psi(f_1\otimes\dots\otimes f_n)\}\end{align}
with the differential
\begin{align}d\psi(f_0\otimes \dots \otimes f_n)=\sum_{i\in \mathbb Z/(n+1)\mathbb Z} (-1)^{in}\psi(f_if_{i+1}\otimes f_{i+2}\otimes \dots \otimes f_{i-1})\,.\end{align}
The existence of $\eta$ satisfying the constraint \eqref{cocycle} means that the 2-cocycle $\Psi$ is a coboundary. The obstruction lies in $H^2_{cycl}(\A)$.

In the quiver case for $\A=\C^{Q_0}$, there is no obstructions as $H^2_{cycl}(\C^{Q_0})=0$. In the manifold case, the $2$-nd continuous cyclic cohomology of $\A=C^\infty_\C(X)$ coincides with the continuous dual to $\Omega^1(X)/d \Omega^0(X)$. Assume for simplicity that $X$ is oriented. In this case, a dense subset of the continuous dual, as above, consists of closed forms on $X$ of degree equal to $\dim(X)-1$. Any closed form
\begin{align} \beta \in \Gamma(X,\wedge^{\dim X -1} T^*X\otimes_\R C),\quad d\beta=0\end{align}
 gives a cyclic 2-cochain by the formula 
\begin{align}f_1\otimes f_2\mapsto \int_X f_1 df_2\wedge \beta\,. \end{align}
In our example of a complex distribution $\mathcal F\subset TX\otimes \C$ and a Hermitian form on $\mathcal F$, the corresponding obstruction class in  $H^2_{cycl}(\A)$ is represented by the differential of certain form $\delta$ of degree $\dim_\R X-2$. The vanishing of the obstruction means that $\delta$ is closed. This is a necessary and sufficient condition for the existence of a solution $\eta$ for the constraint  \eqref{cocycle}. In the case of HYM equations on complex  K\"{a}hler manifolds the form $\delta$ is equal to $(\omega_X)^{\dim_\C X-1}$ where $\omega_X$ is the K\"{a}hler form on $X$.
\begin{remark}
We already observed that (for a given data $\bf{A1},\bf{A2},\bf{A3},\bf{A4}$) the obstruction to the existence of functional $\eta$ is a class in $H^2_{cycl}(\A)$ satisfying the reality constraint \eqref{cocycle-reality}.
 If the obstruction vanishes, the set of choices of possible functionals $\eta$ is a torsor over the real subspace of  $H^1_{cycl}(\A)=Hom(A/[A,A],\C)$ given by the fixed points of the anti-linear involution 
 \begin{align}\eta\mapsto \eta^\sigma,\quad \eta^\sigma(f):=-\cc{\eta(f^*)}\,.
 \end{align}
 Notice the similarity with the analogous  question for the liftings of a symplectic action to a Hamiltonian one, cf. Remark \ref{h1h2}.
\end{remark}

\

{\bf(M3)}: In the quiver case, a Hermitian  $\A$-valued form on $\A$-module $E$ is equivalent to the collection of Hermitian forms on the individual complex vector spaces $\E_v$ for all vertices $v\in Q_0$.

 In the manifold case (independently on the choice of complex distribution $\mathcal F$), a Hermitian $\A$-valued form on an $\A$-module $E=\Gamma(X,\mathcal E)$ is equivalent to a Hermitian norm on the  corresponding complex vector bundle $\mathcal E$.
 
 In general, when a projector $P\in Mat(n\times n,\A),\,P^2=P $ is \emph{self-adjoint}:
 \begin{align}\label{selfadjoint}
    P=(p_{ij})_{1\le i,j\le n}\in Mat(n\times n, \A),\quad p_{ij}^*=p_{ji}\,\,\forall i,j,\quad P^2=P\,,
\end{align}
then the  submodule $E:=\A^n\cdot P$ carries an $\A$-valued Hermitian form given by the restriction to $E\subset \A^n$ of the standard form on $\A^n$:
\begin{equation}\label{standradform}
\mathsf H_{\operatorname{standard}}((f_1,\dots,f_n),(\cc{g_1},\dots,\cc{g_n})):=\sum_i f_i g_i^*\,.
\end{equation}

\begin{remark}
The framework of [\ref{connes}] (and then of [\ref{schwarz}]) fits (partially) into our setup. In order to define the  notion of a connection, 
 authors of [\ref{connes}] use a collection $(\partial_i)_{i=1,\dots n}$ of derivations of an algebra $\mathcal A$ closed under the Lie bracket. In our formalism the corresponding bimodule is $\Omega^1:=\A_{diag}^{\oplus n}$
  endowed with the derivation
  \begin{align}
      d(f):=(\partial_1 f,\dots,\partial_n f)\in \Omega^1\,.
  \end{align}
\end{remark}

\ 

\section{Formula for the Hamiltonian action}

\subsection{The case of a trivial bundle}

\

Let us assume that $E\simeq \A^n=\C^n\otimes \A$ is a free finitely generated \emph{left} module over $\A$, endowed with the canonical Hermitian $\A$-valued form (see \eqref{standradform}).

The set $M$ of connections on $E$ can be identified in the usual way with the space of matrices of 1-forms:

\begin{equation} A=(A_{ij})_{1\le i,j,\le n}\in Mat(n\times n,\Omega^1)\quad  \leftrightsquigarrow \quad{\nabla_A}: E\to \Omega^1\otimes_\A E,\quad \nabla_A(\phi)=d\phi+\phi\cdot A\,.
\end{equation}

The Lie algebra of the ``compact  gauge group'' is defined as 
 \begin{equation}\label{unitary}
\mathfrak{k}:=\{(u_{ij})_{1\le i,j \le n}\in Mat(n\times n,\A)\,|\, u_{ij}^*=-u_{ji} \quad \forall i,j\}\,.
\end{equation}
It acts on the (infinite-dimensional) complex affine space $M$ of connections by the infinitesimal affine transformations
 \begin{equation}(d+\cdot A)\mapsto (1-\cdot \epsilon u)\circ (d+\cdot A)\circ (1-\cdot \epsilon u)^{-1}=d+\cdot A+\cdot\epsilon(du+[u,A])
\end{equation}
where $\epsilon$ is a formal variable satisfying $\epsilon^2=0$, and notation $\cdot A$ stays for the operator of \emph{right} multiplication by  $A$, and similarly for other symbols. In other words, the value of the  vector field $X_u$ corresponding to $u\in \mathfrak{k}$ on $M$ at the point  $A$ is 
\begin{align} \label{formulaX}
    {X_u}_{|A}=du+[u,A]\,.
\end{align}

\

An $\A$-valued Hermitian form $\mathsf H_0$ on $E$ together with a ``noncommutative K\"{a}hler metric" $\omega$ produces a usual $\C$-valued Hermitian form
 on the complex vector space $ Mat(n\times n,\C)\otimes \Omega^1$ given by
 \begin{equation}\label{omega0} \omega_0(A^{(1)},\cc{A^{(2)}}):= \sum_{ij} \omega(A^{(1)}_{ij},\cc{A^{(2)}_{ij}})\,.
\end{equation}
This form is strictly positive on non-zero vectors by \eqref{kahlerpositive}, and the infinitesimal action of $\mathfrak{k}$ via $A\mapsto A+\epsilon[u,A]$ preserves $\omega_0$. Therefore, the infinitesimal action of $\mathfrak{k}$ on affine space $M$ of connections 
endowed with the ``constant" K\"{a}hler metric corresponding to $\omega_0$ is by \emph{K\"{a}hler isometries}, because the vector field $X_u$ is the sum of the infinitesimal generator of the linear action $A\mapsto A+\epsilon[u,A]$ (which is an isometry), and of the shift by a constant vector $A\mapsto A+\epsilon\,du$ (which is also an isometry). 

In what follows, we will use an identity which follows directly from \eqref{kahlerpositive} and the definition \eqref{omega0}
\begin{align}\label{leibniz}
    \omega_0([A^{(1)},u],\cc{A^{(2)}})=\omega_0(A^{(1)},\cc{[u,A^{(2)}]})\quad \forall A^{(1)},A^{(2)}\in M,\,\,\forall u\in \mathfrak k\,.
\end{align}

\ 

The constant (i.e. invariant under shifts)  symplectic form $\omega_M^{symp}$ on $M$ corresponding to the K\"{a}hler metric  $\omega_0$ is given by the real skew-symmetric form on the tangent space
 \begin{equation}\label{defomega}
 \omega_M^{symp}(A^{(1)},A^{(2)}):=\im \omega_0(A^{(1)},\cc{A^{(2)}})=\frac{1}{2\sqrt{-1}}\left(\omega_0(A^{(1)},\cc{A^{(2)}})-
 \omega_0(A^{(2)},\cc{A^{(1)}})\right)\,.
\end{equation}

For a given $u\in \mathfrak{k}$, the corresponding vector field $X_u$ is an infinitesimal K\"{a}hler isometry, hence it preserves the symplectic form $\omega_M^{symp}$. We claim that this symplectic action of $\mathfrak{k}$ can be lifted to a Hamiltionian action. Let us denote for $u\in \mathfrak{k}$ by $H_u$ the following real-valued function on $M$:
\begin{align}\label{momenttrivial}
 H_u(A):=\eta(\text{Trace}(u))-\omega^{symp}_M(A,du)+\frac{1}{2}\omega^{symp}_M(A,[A,u]) \,.  
\end{align}
\begin{prop}
 The assignment $u\mapsto H_u$ is a Lie algebra homomorphism lifting the action $u\mapsto X_u$.
\end{prop}
{\it Proof}: First, it is immediate to see that the vector field $X_u$ corresponds to the Hamiltionian $H_u$:
\begin{align}
    i_{X_u} \omega_M^{symp}=d H_u\,.
\end{align}
It suffices (see \eqref{momentmap2}) to prove that
\begin{align}
    \omega^{symp}_M(X_{u_1},X_{u_2})=H_{[u_1,u_2]}\quad \forall u_1,u_2\in \mathfrak{k}\,.
\end{align}

\ 

In other words, we have to check that for any $A\in M$

\begin{multline}\label{bracketu1u2}
     \omega^{symp}_M(du_1+[u_1,A],du_2+[u_2,A])=\\
    =\eta(\text{Trace}([u_1,u_2]))-\omega^{symp}_M(A,d[u_1,u_2])+\frac{1}{2}\omega^{symp}_M(A,[A,[u_1,u_2]])\,.
\end{multline}

Indeed, we have
\begin{multline}
    \omega^{symp}_M(du_1+[u_1,A],du_2+[u_2,A])\stackrel{\eqref{defomega}}{=}\im\omega_0\left(du_1+[u_1,A],\cc{du_2+[u_2,A]}\right)=\\
    =\im\omega_0(du_1,\cc{du_2})+\im\omega_0\left([u_1,A],\cc{du_2}\right)+\im\omega_0\left(du_1,\cc{[u_2,A]}\right)+\im\omega_0\left([u_1,A],\cc{[u_2,A]}\right)\,.
\end{multline}
Then we use
\begin{multline}
    \im\omega_0(du_1,\cc{du_2})=\frac{1}{2\sqrt{-1}}\left(\omega_0(du_1,\cc{du_2})-\omega_0(du_2,\cc{du_1})\right)=
    \\
    \stackrel{\eqref{omega0}}{=}\frac{1}{2\sqrt{-1}}\sum_{ij}\left( \omega(d(u_1)_{ij},\cc{d(u_2)_{ij}} )
    -\omega(d(u_2)_{ij},\cc{d(u_1)_{ij}} )\right)=\\
    \stackrel{\eqref{unitary}}{=} \frac{-1}{2\sqrt{-1}}\sum_{ij}\left( \omega(d(u_1)_{ij},\cc{d(u_2)_{ji}^*} )
    -\omega(d(u_2)_{ij},\cc{d(u_1)_{ji}^*} )\right)=\\
   \stackrel{\eqref{cocycle}}{=} \sum_{ij}\left( \eta([(u_1)_{ij},(u_2)_{ji}])  \right)=\eta(\text{Trace}([u_1,u_2]))\,,\\
    \end{multline}
    \begin{multline}
    \im\omega_0\left([u_1,A],\cc{du_2}\right)+\im\omega_0\left(du_1,\cc{[u_2,A]}\right)=\\
    =
    -\im\omega_0\left([A,u_1],\cc{du_2}\right)+\im\omega_0\left([A,u_2],\cc{du_1}\right)\stackrel{\eqref{leibniz}}{=}
     -\im\omega_0\left(A,\cc{d[u_1,u_2]}\right)=-\omega_M^{symp}(A,d[u_1,u_2])\,,\\
\end{multline}
and, utilizing the antisymmetry of $\omega^{symp}_M$,
\begin{multline}\label{double}
    \im\omega_0\left([u_1,A],\cc{[u_2,A]}\right)=\frac{1}{2}\left(\im\omega_0\left([A,u_1],\cc{[A,u_2]}\right)-\im\omega_0\left([A,u_2],\cc{[A,u_1]}\right)\right)=\\
   \stackrel{\eqref{leibniz}}{=} \frac{1}{2} \im\omega_0\left(A,\cc{[u_1,[A,u_2]]-[u_2,[A,u_1]]}\right)=
   \frac{1}{2} \im\omega_0\left(A,\cc{[A,[u_1,u_2]]}\right)=\frac{1}{2}\omega^{symp}_M(A,[A,[u_1,u_2]])\,.
\end{multline}

\ 

This calculation finishes the proof of \eqref{bracketu1u2}.
 $\blacksquare$
 
 \ 
 
 Later we will need a formula for $H_u(A)$ written in a slightly different form:
 \begin{multline}\label{momenttrivial2}
   H_u(A)\stackrel{\eqref{momenttrivial}}{=}  \eta(\text{Trace}(u))-\omega^{symp}(A,du)+\frac{1}{2}\omega^{symp} (A,[A,u])=\\
   \stackrel{\eqref{defomega}}{=}\eta(\text{Trace}(u))+\frac{1}{2\sqrt{-1}}\Bigl[-\omega_0(A,\cc{du})+\omega_0(du,\cc A)+\frac{1}{2}\omega_0(A,\cc{[A,u]})-\frac{1}{2}\omega_0([A,u],\cc A)\Bigr]=\\
   \stackrel{\eqref{leibniz}}{=}\eta(\text{Trace}(u))+\frac{1}{2\sqrt{-1}}\Bigl[-\omega_0(A,\cc{du})+\omega_0(du,\cc A)-\omega_0([A,u],\cc A)\Bigr]\,.
 \end{multline}

\

\subsection{General bundle}

\

\

Let $P$ be a self-adjoint (see \eqref{selfadjoint}) projector in $Mat(n\times n,\A)$. 
Then the free module $E=\A^n$ splits into the orthogonal sum of two submodules (here we denote $\id_{\A^N}$ as $1$ for brevity)
\begin{align}\label{decomp}
    E\simeq E_1\oplus E_2,\quad E_1:=E\cdot P,\,\,E_2:=E\cdot (1-P)\,.
\end{align}
We will consider the action of the gauge group of unitary automorphisms of $E_1$ on the space $M_1$ of connections on $E_1$. First, consider the  Lie subalgebra $\mathfrak  k_{1+2}$ of $\mathfrak k$ consisting  of infinitesimal unitary symmetries preserving the direct sum decomposition \eqref{decomp}
\begin{align}
    \mathfrak k_{1+2}:=\{u\in \mathfrak k\,|\, u=PuP+(1-P)u(1-P)\}\,.
\end{align}
It is clear that $\mathfrak k_{1+2}$ is the direct sum of two subalgebras
\begin{align}\label{unitaryproj}
    \mathfrak k_{1}:=\{u\in \mathfrak k\,|\, u=PuP\},\quad \mathfrak k_{2}:=\{u\in \mathfrak k\,|\, u=(1-P)u(1-P)\}
\end{align}
and $\mathfrak k_1$ is the  Lie algebra of infinitesimal unitary symmetries of $E_1$.

\ 

Next, consider the  space of connections on $E$ preserving the direct sum decomposition \eqref{decomp}:
\begin{align}
    \mathfrak M_{1+2}:=\{A\in Mat(n\times n, \Omega^1)\,|\, d+A=P\cdot (d+A)\cdot P+(1-P)\cdot (d+A)\cdot (1-P)\}\,.
\end{align}
It is an affine subspace of the affine space $M$ of connections on $E$, and it  is isomorphic to the product of the space $M_1$ of connections in $E_1$ and the space $M_2$ of connections in $E_2$.

\ 

There is a distinguished point $A_{can}\in M_{1+2}$ given by
\begin{align}
    A_{can}=P\cdot dP+(1-P)\cdot d(1-P)=(2P-1)\cdot dP
\end{align}
which gives points $A_{can,1}\in M_1,\,\, A_{can,2}\in M_2$ after the  identification $M_{1=2}\simeq M_1\times M_2$.
Then we identify  $M_1$ with an affine  \emph{subspace} $M_{(1)}\subset M_{1+2}$ consisting of connections whose restriction to $E_2$ is
$A_{can,2}$. Explicitly, we have
\begin{align}
    M_{(1)}=
\{A\in M\,|\, A=A_{can}+\delta_A,\,\,\,\delta_A=P\delta_A P\}\,.
\end{align}
The Lie subalgebra $\mathfrak k_1\subset \mathfrak k$ preserves the submanifold $M_{(1)}\subset M$. In particular, for any $u\in \mathfrak k_1$ the value of the vector field $X_u$ restricted to $M_{(1)}$ is given (see \eqref{formulaX}) at the point $A_{can}+\delta_A$ by
\begin{align}
   {X_u}_{|A_{can}+\delta_A}=du+[u,A_{can}+\delta_A] \,.
\end{align}
Using \eqref{bracketu1u2}, this formula implies for any $u_1,u_2\in \mathfrak k_1$
\begin{multline}
\omega^{symp}(X_{u_1},X_{u_2})=
     \omega^{symp}(du_1+[u_1,A],du_2+[u_2,A])=\\
    =\eta(\text{Trace}([u_1,u_2]))-\omega^{symp}(A,d[u_1,u_2])+\frac{1}{2}\omega^{symp}(A,[A,[u_1,u_2]])\,,
\end{multline}
where $A:=A_{can}+\delta_A$. We conclude 
\begin{prop}\label{propaction} The assignment
\begin{multline}\label{momentproj}
H_{(1),u}(A_{can}+\delta_A):=\\
=\eta(\operatorname{Trace}(u))-\omega^{symp}(A_{can}+\delta_A,du)+\frac{1}{2}\omega^{symp}(A_{can}+\delta_A,[A_{can}+\delta_A,u])
\end{multline}
gives a Hamiltionian action of $\mathfrak k_1$ on $M_{(1)}\simeq M_1$ lifting the symplectic action by gauge transformations. $\blacksquare$
\end{prop}

\ 

\subsection{Universal formula for the moment map}\label{sectionuniversal}

\ 

\

In this section we  propose a formula for the moment map written in an ``invariant" way, which \emph{does not} refer explicitly to the representation of finitely-generated  projective $\A$-module $E$ as an image of a self-adjoint projector  $P\in Mat(n\times n, \A)$ for some $n<\infty$. 

In order to be able to write the formula, we will need to introduce some notations and constructions.

\ 
\subsubsection{More about Hermitian modules} In this section $\A$ denotes an arbitrary  $*$-algebra.  Recall (see \eqref{hermiso}) that a Hermitian structure $\mathsf H$ on a finitely-generated $\A$-module $E$ gives rise to an isomorphism of $\A^{op}$-modules
\begin{align}\label{hermiso1} iso_{\,\mathsf H}:\cc{E}\simeq E^{\vee}\end{align}
(here we consider the $\cc{A}$-module $\cc{E}$ as an  $\A^{op}$-module via the canonical isomorphism of algebras $\cc{A}\simeq \A^{op} $).

With any endomorphism 
$$ u:E\to E,\quad u\in Hom_{A-mod}(E,E) $$
we can associate
\begin{enumerate}
    \item the {\it complex-conjugate} morphism 
    $$\cc{u}:\cc{E}\to \cc{E},\quad \cc{u}\in Hom_{\cc{A}-mod}(\cc{E},\cc{E}) \,,$$
\item the {\it adjoint} morphism, by applying the {\it contravariant} functor $Hom_{\A-mod}(\cdot,\A)$
 $$u^t:E^\vee\to E^\vee, \quad u^t \in Hom_{mod-\A}(E^\vee, E^\vee)\,.$$
\end{enumerate}
The Lie algebra of infinitesimal unitary symmetries of $(E,\mathsf H)$ is defined (generalizing \eqref{unitary},\eqref{unitaryproj}) as
\begin{align}\label{unitaryuniv}
    \mathfrak k:=\{u\in Hom_{\A-mod}(E,E)\,|\, u^\dagger=-u\}, \quad u^\dagger:=\cc{u}^t=\cc{u^t}\,.
\end{align}

\ 

Recall (see \eqref{bimoduleB},\eqref{bimoduleBmor} and \eqref{nablaprim},\eqref{nablaprimconstraint}) that a connection $\nabla$ on $\A$-module $E$ we can recast as a homomorphism of $\A$-modules 
 \begin{align}
     \wn\in Hom_{\A-mod}(E,B\otimes_\A E)
 \end{align}
 such that 
 \begin{align}(\pi_B\otimes \id_E)\circ \wn=\id_E \,. \end{align}
 The complex conjugation gives
 \begin{align}
     \cc{\wn}\in Hom_{\cc{\A}-mod}(\cc{E},\cc{B}\otimes_{\cc{\A}} \cc{E}),\quad (\cc{\pi_B}\otimes \id_{\cc{E}})\circ \cc{\wn}=\id_{\cc{E}} \,. 
 \end{align}
 Applying the isomorphism $iso_{\mathsf H}$ from \eqref{hermiso1} we obtain another morphism
 \begin{align}
     \widehat{\nabla}\in Hom_{mod-\A}(E^\vee, E^\vee\otimes_\A \cc{B}),\quad ( \id_{E^\vee}\otimes \cc{\pi_B})\circ      \widehat{\nabla}=\id_{E^{\vee}} \,,
 \end{align}
 where we treat $\cc{B}$ as a $\A\otimes \A^{op}$-module via the canonical isomorphism of algebras $\A\otimes \A^{op}\simeq \cc{\A}\otimes \cc{\A^{op}}$. Finally, using the following chain of isomorphisms
 \begin{multline}\label{chain1}
     Hom_{mod-\A}(E^\vee, E^\vee\otimes_\A \cc{B})\simeq Hom_{mod-\A}(E^\vee, Hom_{\A-mod}  (E,\cc{B}))\simeq\\\simeq Hom_{\A-mod-\A}(E\otimes_\C E^\vee,\cc{B})\simeq Hom_{\A-mod}(E,Hom_{mod-\A}(E^\vee,\cc{B}))\simeq Hom_{\A-mod}(E,\cc{B}\otimes_\A E)
 \end{multline}
 we obtain the {\it Hermitian-conjugate} connection (formulated in terms of a morphism of $\A$-modules)
 \begin{align} 
 {\wn^\dagger}\in Hom_{\A-mod}(E,\cc{B}\otimes_\A E), \quad (\pi_{\cc{B}}\otimes \id_E)\circ {\wn^\dagger}=\id_E\,.
 \end{align}
  Here in \eqref{chain1} we use the fact that for any finitely-generated projective $\A$-module $E$ and an {\it arbitrary} $\A$-module $F$, the canonical map
  \begin{multline}\label{adjunction}
      E^\vee \otimes_\A F\to Hom_{\A-mod}(E ,F),\\ E^\vee\otimes_\A F=Hom_{\A-mod}(E,\A)\otimes_\A Hom_{\A-mod}(\A,F)\xrightarrow{\rm {composition}}Hom_{\A-mod}(E,F)
  \end{multline}
  is an isomorphism.
  
  \

Alternatively, let us use the  {\it Hermitian-conjugate} derivation $d^\dagger$  with values in $\cc{\Omega^1}$ (see \eqref{daggerderivation})
\begin{align}\label{daggerder}
    d^\dagger: \A\to \cc{\Omega^1},\quad d^\dagger(f):=-\cc{d(f^*)},\qquad d^\dagger(f\cdot g)=f\cdot d^\dagger(g)+d^\dagger(f)\cdot g\,\text{ for free}\,.
\end{align}
 The bimodule $\cc{B}$ is identified with
 \begin{align}
 \A\oplus \cc{\Omega^1}, \text{ with the bimodule structure }    f\cdot(h,\cc{\alpha})\cdot g:=
 (f\cdot h\cdot g, f\cdot \alpha\cdot g +d^\dagger f\cdot h\cdot g)\text{ (as in \eqref{bimoduleB})}
 \end{align}
 by the map
 \begin{align}
     \cc{(h,\alpha)}\in \cc{B}\mapsto (h^*,\cc{\alpha}-\cc{dh})\in \A\oplus \cc{\Omega^1}\,.
 \end{align}
 For the trivial $\A$-module $E=\A^n$ with the canonical Hermitian $\A$-valued form \eqref{standradform}, for any connection $A$ given by a $(n\times n)$ matrix 
 \begin{align}
     A=(A_{ij})_{1\le i,j\le n}\in Mat(n\times n,\Omega^1)
 \end{align}
 the Hermitian conjugate connection is given by
 \begin{align}
     A^\dagger=((A^\dagger)_{ij})_{1\le i,j\le n}\in Mat(n\times n,\cc{\Omega^1}),\qquad (A^\dagger)_{ij}:=\cc{A_{ji}}\quad\forall \, i,j\,.
 \end{align}
 
 \ 
 
 \subsubsection{Bimodules and traces} In this section $A$ denotes an arbitrary associative algebra over $\C$ (not necessarily a $*$-algebra).    With every $\A$-bimodule $G$ we associate a vector space $\# (G)$ by the formula
 \begin{align}
     \#(G):=G/\{\text{linear span of }a\cdot g-g\cdot a\,| \, a\in \A, g\in G\}\simeq G\otimes_{\A\otimes \A^{op}} \A_{diag}\,.
 \end{align}
 It follows form the definition that for any finite sequence of bimodules $G_1,\dots, G_n)$ one has a chain of canonical isomorphisms
 \begin{align}\label{dashcyclic}
     \#(G_1\otimes_\A G_2\otimes_A\dots\otimes_\A G_n)\simeq \#(G_2\otimes_\A G_3\otimes_A\dots\otimes_\A G_1)\simeq \#(G_n\otimes_\A G_1\otimes_A\dots\otimes_\A G_{n-1})\,.
 \end{align}
 
 \ 
 
 For any finitely-generated $\A$-module $E$, any $\A$-bimodule $G$ and any morphism of $\A$-modules
 \begin{align}
     \Phi: E\to G\otimes_\A E
 \end{align}
 we define its {\it trace along }$E$ (denoted by $\operatorname{Trace}_E(\Phi)$) with values in $\#(G)$, via the chain of isomorphisms
 \begin{align}
     \Phi\in Hom_{\A-mod}(E,G\otimes_\A E)\stackrel{\eqref{adjunction}}{\simeq}E^\vee\otimes_\A G\otimes_A E\simeq G\otimes_{\A\otimes \A^{op}} (E\otimes_\C E^\vee)
 \end{align}
 and a map
 \begin{align}
    G\otimes_{\A\otimes \A^{op}} (E\otimes_\C E^\vee) \xrightarrow{\id_G\otimes_{\A\otimes A^{op}} \delta_E}  G\otimes_{\A\otimes \A^{op}} \A_{diag}=\#(G)\ni \operatorname{Trace}_E(\Phi)\,,
 \end{align}
 where 
 \begin{align}
     \delta_E: E\otimes_\C E^\vee \to \A_{diag},\quad \delta_E (e\otimes e^\vee):=e^\vee(e)\in \A,\,\,\forall e\in E,\, \forall e^\vee\in Hom_{\A-mod}(E,A)=E^\vee
 \end{align}
 is the canonical morphism of $\A$-bimodules.
 
\begin{remark}
The constraints for the left and right action form on the noncommutative K\"{a}hler form $\omega$ (see \eqref{kahlerpositive})   can be interpreted as follows: $\omega$ is equal to the composition of a linear functional
\begin{align}
    \omega':\#(\Omega^1\otimes_\A \cc{\Omega^1})\to \C\
\end{align}
 and of the canonical surjection 
 \begin{align}\label{omegaprime}
   \Omega^1\otimes_\C \cc{\Omega^1} \twoheadrightarrow \#(\Omega^1\otimes_\A \cc{\Omega^1})\,.
\end{align}
\end{remark}
 \ 
 
 \subsubsection{Linear functional on the triple tensor product} In this section we work in the 
  setup $\bf (A1)\rm{-}\bf (A5)$. Recall that we have automatically \emph{two} derivations $d,d^\dagger$ (see \eqref{daggerder}), hence we can define a \emph{doubled} bimodule by
  \begin{align}
      \mathbb B:=\A\oplus \Omega^1\oplus \cc{\Omega^1}, \quad f\cdot(h,\alpha,\cc{\beta})\cdot g:=(f\cdot h\cdot g,f\cdot \alpha \cdot g+ df \cdot h\cdot g,f\cdot \cc{\beta}\cdot g+d^\dagger f\cdot h\cdot g)\,.
  \end{align}
  Define a linear map 
  \begin{align}
      \Xi:\mathbb B\otimes_\C\mathbb B \otimes_\C \mathbb B\to \C
  \end{align}
 by the following formulas (the missing terms map to zero):

 \begin{align}\label{xi1}
     f_1\otimes f_2\otimes f_3 &\mapsto \frac{1}{3}\left(-2\sqrt{-1}\,\eta(f_1 f_2 f_3)-\omega(df_1, d^\dagger f_2\cdot  f_3)+\omega( df_2 \cdot f_3, d^\dagger f_1)\right)+(1\to 2\to 3)\,,\\
     \label{xi2}
   \alpha_1\otimes f_2\otimes f_3 &\mapsto +\omega(\alpha_1, d^\dagger f_2\cdot f_3),\qquad \text{ and the same r.h.s. for } f_3\otimes  \alpha_1\otimes f_2, \,\,f_2\otimes f_3\otimes \alpha_1\,,\\
   \label{xi3}
   \cc{\alpha_1}\otimes f_2\otimes f_3 &\mapsto -\omega( d f_2\cdot f_3, \cc{\alpha_1})
  \qquad\,\,\,\,\,\,\text{ and the same r.h.s. for } f_3\otimes \cc{\alpha_1}\otimes f_2, \,\,f_2\otimes f_3\otimes \cc{\alpha_1}\,, \\
  \label{xi4}
   \alpha_1\otimes \cc{\alpha_2}\otimes f_3&\mapsto -\omega(\alpha_1, \cc{\alpha_2}\cdot f_3)
  \qquad\,\,\,\,\,\,\,\text{ and the same r.h.s. for } f_3\otimes \alpha_1\otimes \cc{\alpha_2}, \,\, \cc{\alpha_2}\otimes f_3 \otimes \alpha_1\,, \\
  \label{xi5}
   \cc{\alpha_1}\otimes {\alpha_2}\otimes f_3&\mapsto +\omega( {\alpha_2}\cdot f_3, \cc{\alpha_1})
   \qquad\,\,\,\,\,\,\,\,\text{ and the same r.h.s. for } f_3\otimes \cc{\alpha_1}\otimes \alpha_2, \,\,\alpha_2\otimes f_3\otimes \cc{\alpha_1}\,.
 \end{align}

 \begin{prop}\label{Xiprop}
 The map $\Xi$ descends to a map 
 \begin{align}
     \Xi': \#(\mathbb B \otimes_\A\mathbb B\otimes_\A\mathbb B)\to \C\,.
 \end{align}
 \end{prop}
 {\it Proof}: It follows from the definition that the map $\Xi$ is $\mathbb Z/3\mathbb Z$-invariant, where $\mathbb Z/3\mathbb Z$ acts\footnote{Notice, that the cyclic group action descends to $\#(\mathbb B \otimes_\A\mathbb B\otimes_A\mathbb B)$ by \eqref{dashcyclic}. The functional $\Xi'$ is cyclically invariant as well.} by cyclic permutations of factors in $\mathbb B^{\otimes 3}$. This symmetry reduces  the number of possible checks to the following list:
 
 \begin{align}\label{nontrivialcheck}
     -f_1\d g\o f_2\o f_3+f_1\o g\d f_2\o f_3+f_1\o dg\d f_2\o f_3+ f_1\o d^\dagger g \d f_2\o f_3 &\stackrel{?}{\mapsto} 0\,,\\
     -f_1\d g\o f_2\o \alpha_3 + f_1\o g\cdot f_2\o \alpha_3+f_1\o d^\dagger g\d f\o \alpha_3&\stackrel{?}{\mapsto} 0\,,\\
    -f_1\d g\o f_2\o \cc{\alpha_3} + f_1\o g\cdot f_2\o \cc{\alpha_3}+f_1\o dg\d f\o \cc{\alpha_3}&\stackrel{?}{\mapsto} 0\,,\\
     -f_1\cdot g \o \alpha_2\o f_3+f_1\o g\d \alpha_2 \o f_3 &\stackrel{?}{\mapsto} 0\,,\\
      -f_1\cdot g \o \cc{\alpha_2}\o f_3+f_1\o g\d \cc{\alpha_2} \o f_3 &\stackrel{?}{\mapsto} 0\,,\\
      -\alpha_1\d g\o f_2\o f_3+\alpha_1\o g\d f_2\o f_3+\alpha_1 \o d^\dagger g \d f_2\o f_3  &\stackrel{?}{\mapsto} 0\,,\\
      -\cc{\alpha_1}\d g\o f_2\o f_3+\cc{\alpha_1}\o g\d f_2\o f_3+\cc{\alpha_1} \o d g \d f_2\o f_3  &\stackrel{?}{\mapsto} 0\,.
 \end{align}
 
 \ 
 
 All the checks are straightforward corollaries of the Leibniz rule, of  the fact that $\omega$ descends to a functional $\omega': \#(\Omega^1\otimes_\A \cc{\Omega_1})\to \C$ (see \eqref{omegaprime}), and of the relation \eqref{cocycle}. Here is the most non-trivial check \eqref{nontrivialcheck}. 
 \begin{multline}
 -f_1\d g\o f_2\o f_3+f_1\o g\d f_2\o f_3+f_1\o dg\d f_2\o f_3+ f_1\o d^\dagger g \d f_2\o f_3\mapsto \\   
+\frac{2\sqrt{-1}}{3}\,\Bigl(\eta(g f_2 f_3 f_1 -f_2 f_3 f_1 g)\Bigr)+\\
+\frac{1}{3}\cdot\Bigl(\omega(\underbrace{f_1\cdot dg},d^\dagger f_2 \cdot f_3)+\omega( df_3,\underbrace{f_1\cdot d^\dagger g} \cdot f_2)-\omega(\underbrace{dg\cdot f_2},d^\dagger f_3\cdot f_1)-\omega(df_1,\underbrace{d^\dagger g\cdot f_2}\cdot f_3)-\\
-\omega(\underbrace{f_1\cdot dg}\cdot f_2,d^\dagger f_3)-\omega(df_2\cdot f_3,\underbrace{f_1\cdot  d^\dagger  g})+\omega(\underbrace{dg\cdot f_2}\cdot f_3,d^\dagger  f_1)+\omega(df_3\cdot f_1,\underbrace{d^\dagger g \cdot f_2})\Bigr)+\\
+\omega(dg\cdot f_2, d^\dagger f_3\cdot f_1)-\omega(df_3\cdot f_1, d^\dagger g \cdot f_2)=\\
\stackrel{\eqref{cocycle},\eqref{daggerder}}{=} \frac{1}{3}\Bigl(\omega(dg,d^\dagger(f_2 f_3 f_4))-\omega(d(f_2 f_3 f_4),d^\dagger g)\Bigr)+\\
+\frac{1}{3}\Bigl( -\omega(dg, d^\dagger f_2\cdot f_3 f_1)-\omega(dg, f_2\cdot d^\dagger f_3 \cdot f_1)-\omega(dg, f_2 f_3 \cdot d^\dagger f_1)+ \\
+\omega(df_2\cdot f_3 f_1,d^\dagger g )+\omega(f_2\cdot df_3 \cdot f_1,d^\dagger g )+\omega(f_2 f_3 \cdot df_1,d^\dagger g ) \Bigr)=0\,.
 \end{multline}
 The rest is a routine calculation. 
 $\blacksquare$

\subsubsection{Formula for the moment map in terms of $\Xi'$}
 
Our goal (in the setup $\bf (A1)\rm{-}\bf (A5)$) is to  associate with any Herimitan module $(E,\mathsf H)$ endowed with a connection $\nabla$, a $\R$-linear functional on the Lie algebra $\mathfrak k$ defined as in \eqref{unitaryuniv}. In other words, we want to define a number
\begin{align}\label{momentuni}
    H_u(\mathsf H, \nabla)\in \R
\end{align}
depending $\R$-linearly on $u\in \mathfrak k$, extending the formulas \eqref{momenttrivial},\eqref{momentproj}.

We can form the following a chain of morphisms of $\A$-modules:
\begin{align}\label{chain}
    E\stackrel{u}{\to}E \xrightarrow{\wnn} \mathbb B \otimes_A E
    \xrightarrow{\id_{\mathbb B}\otimes_A \wnn}\mathbb B \otimes_A \mathbb B \otimes_A E\xrightarrow{(\id_{\mathbb B\otimes_\A \mathbb B})\otimes_\A {\wnn}}
    \mathbb B \otimes_A \mathbb B \otimes_A \mathbb B \otimes_A E
\end{align}
where 
\begin{align}
    \wnn: E\to \mathbb B \otimes_\A E
\end{align}
is the morphism of $\A$-modules associated with the connection $\nabla \oplus \nabla^\dagger: E\to (\Omega^1\oplus\cc{\Omega^1})\otimes_\A E$.

The composition in \eqref{chain} is a morphism of $\A$-modules 
\begin{align}
    C_3: E\to \mathbb B \otimes_A \mathbb B \otimes_A \mathbb B \otimes_A E \,.
\end{align}
Applying the trace along $E$ to the morphism $C_3$ we obtain an element
\begin{align}
    \operatorname{Trace}_E(C_3)\in \#(\mathbb B \otimes_A \mathbb B \otimes_A \mathbb B)\,.
\end{align}

\begin{definition}
The moment map (see \eqref{momentuni})  is given by 
\begin{align}\label{mmviaxi}
  H_u(\mathsf H, \nabla):= \frac{\sqrt{-1}}{2}\cdot\Xi'( \operatorname{Trace}_E(C_3))  
\end{align}
where $\Xi'$ is well-defined by Proposition \ref{Xiprop}. The equation on the Hermitian form
\begin{align}\label{universalmmequation}
    H_u(\mathsf H, \nabla)=0 \quad \forall u \in \mathfrak{k}
\end{align} 
we call the {\bf universal moment map equation}.
\end{definition}

\

\begin{prop}
In the case when Hermitian finitely-generated  projective $\A$-module $E$ is isomorphic to the  image of a self-adjoint projector  $P\in Mat(n\times n, \A)$ for some $n<\infty$ endowed with the induced Hermitian $\A$-valued pairing, the definition of the moment map via \eqref{mmviaxi} and as in \eqref{momentproj} agree.
\end{prop}

{\it Proof}: In order to alleviate the notations we will perform the check  in the simplest case when $E$ is the free module of rank $1$ endowed with the standard Hermitian form. Hence, $u$ is a $(1\times 1)$-matrix, which is just an element of $\A$ satisfying $u^*=-u$. Similarly, the connection is an element $\alpha\in \Omega^1$.

The morphism $\wnn:E\to \mathbb B \otimes_\A E$ is given (on the base element $1\in E=\A$) by 
\begin{align}
    \wnn: 1\mapsto (1,\alpha,\cc{\alpha})\otimes 1\,.
\end{align}

The chain \eqref{chain} applied to the element $1\in \A=E$ is given by
\begin{multline}
    1\mapsto u\mapsto (u,u\cdot \alpha,u\cdot \cc{\alpha})\otimes 1\mapsto (u,u\cdot \alpha,u\cdot \cc{\alpha})\otimes (1,\alpha,\cc{\alpha})\otimes 1\mapsto \\
    \mapsto (u,u\cdot \alpha,u\cdot \cc{\alpha})\otimes (1,\alpha,\cc{\alpha})\otimes (1,\alpha,\cc{\alpha})\otimes 1\,.
\end{multline}
Hence, we have to calculate
\begin{align}
    \Xi\bigl( (u,u\cdot \alpha,u\cdot \cc{\alpha})\otimes (1,\alpha,\cc{\alpha})\otimes (1,\alpha,\cc{\alpha})  \bigr)
\end{align}

Term \eqref{xi1} gives
\begin{align}\label{xir1}
    \frac{3}{3}\,(-2\sqrt{-1})\,\eta(u)=-2\sqrt{-1}\,\eta(u)
\end{align}
as $d 1=d^\dagger 1=0$.

Term \eqref{xi2} gives
\begin{align}\label{xir2}
     +\omega(\alpha,d^\dagger u)\,.
\end{align}

Term \eqref{xi3} gives
\begin{align}\label{xir3}
    -\omega(du, \cc{\alpha})\,.
\end{align}

Term \eqref{xi4} gives
\begin{align}\label{xir4}
    -\omega(u\d \alpha , \cc{\alpha})-\omega(\alpha,\cc{\alpha}\cdot u)-\omega(\alpha,u\d \cc{\alpha})\,.
\end{align}

Term \eqref{xi5} gives
\begin{align}\label{xir5}
+\omega(\alpha,  u\d \cc{\alpha})+\omega(\alpha\d u,\cc{\alpha})+\omega( u\d\alpha,\cc{\alpha})\,.
\end{align}

Taking the sum of \eqref{xir1},\eqref{xir2},\eqref{xir3},\eqref{xir4},\eqref{xir5} we obtain
\begin{align}\Xi'( \operatorname{Trace}_E(C_3))  =
   -2\sqrt{-1}\, \eta(u)+\omega(\alpha,d^\dagger u)-\omega(du, \cc{\alpha})+\omega(\alpha\d u,\cc{\alpha})-
    \omega(u\d \alpha,\cc{\alpha})\,.
\end{align}

Using the fact
\begin{align}
    d^\dagger u:=-\cc{ d u^*}=+\cc{d u}
    \end{align}
 we see that the result is the same as in  \eqref{momenttrivial2}. 
 
 The case of a higher rank trivial bundle, or, more generally, of an image of a Hermitian projector, is completely parallel. $\blacksquare$

\section{Examples} 

\subsection{Quiver type}

The case of a quiver was essentially described above. The algebra $\A$ is $\C^{Q_0}$, the bimodule $\Omega^1$
is $\C^{Q_1}$, the derivation $d$ is 0. The choice of functional $\eta$ corresponds to the choice of a real cyclic 1-cocycle of $\A$.
The resulting moment map equation is thus the general King's equation.

As particular examples relevant for gauge theory we would mention 
ADHM equations \eqref{ADHMeqs}, deformed ADHM equations \eqref{defADHMeqs}, and the 0-dimensional reduction of HYM:  $[z_1,z_2]=0,[z_1^\dagger,z_1]+[z_2^\dagger,z_2]=0$.

\subsection{Manifold type}

For a real Riemannian or for a complex K\"{a}hler manifold $X$ we set $\A:=C^\infty(X)\otimes_\R\C$, the bimodule $\Omega^1$
is either $\Gamma(X,T^*_X\otimes_\R \C)$ or $\Gamma(X,(T^{0,1})^*)$. We get HYM equations in the complex case, and a real version in the totally real case. In the case of \emph{flat} connection over a Riemannian manifold we obtain the well-known equation for the harmonic metric on a non-unitary local system.

In the mixed real/complex case one gets a generalization which coincides with \emph{ Bogomolny equations} when $\dim_\R X=3$ and the complex  distribution
 $\mathcal{F}$ is in local coordinates $(x_1,x_2,x_3)$ generated by 
 \begin{align} \C\cdot \partial_{x_1}+\C\cdot(\partial_{x_2}+i\partial_{x_3})\,.
 \end{align}

\subsection{Mixed manifold/quiver case}
\subsubsection{Twisted quiver bundles (following [\ref{alvarez}])}
Suppose that we are given a K\"{a}hler manifold $X$ with a K\"{a}hler form $\omega_X^{1,1}$, a finite quiver $Q$, and   a collection of holomorphic vector 
bundles $M_a$ over  $X$ for each arrow $a\in Q_1$, endowed with  Hermitian metrics $H_a$.
Then we have the following algebra $\A: = \mathbb{C}^{Q_0} \otimes C^\infty_\C (X)$. The bimodule $\Omega^1$ defined as
\begin{equation}\label{quiverbundle}
    \Omega^1 : = \left(\bigoplus_{v \in Q^0} \pi_v \cdot \Omega^{0,1}(X) \cdot \pi_v \right) \oplus \left(\bigoplus_{a\in Q_1} \pi_{s(a)}\cdot \Gamma( X,  C^\infty_{X,\C}\otimes_{\mathcal O_X}M^{\star}_{a})\cdot \pi_{t(a)}\right).
\end{equation}
 and the derivation $d:\A\to \Omega^1$ is $\cc{\partial}$-operator taking values in the first summand of \eqref{quiverbundle}. 
 
 An example of a module with a connection is an \emph{$M$-twisted $Q$-bundle}, which is by definition (see [\ref{alvarez}]) a  collection of holomorphic vector bundles $(\mathcal E_v)_{v\in Q_0}$ together with a collection of holomorphic morphisms
 \begin{align}
   \forall a\in Q_1:\qquad \phi_a: M_{a}\otimes \mathcal E_{s(a)}\to \mathcal E_{t(a)}\,.
 \end{align}

For such a module an $\A$-valued Hermitian form is a collection of Hermitian metrics $(h_v)_{v\in Q_0}$ on the  individual bundles $\mathcal E_v$.
 Let $\rho$ and $\sigma$ be collections of real numbers $\rho_v$ and $\sigma_v >0$. The harmonicity  equation on $(h_v)_{v\in Q_0}$  (e.g. the moment map equation) is called \emph{twisted quiver $(\rho, \sigma)$-vortex equation}, and it  is:
\begin{equation}
\forall v\in Q_0:\quad 
    \sigma_v \sqrt{-1} \Lambda F_{H_{v}} +  \sum_{a \in s^{-1}(v)} \phi_a \circ \phi^{\star H_a}_a - \sum_{a \in t^{-1}(v)} \phi^{\star H_a}_a \circ \phi_a = \rho_v \, \id_{\mathcal E_v}\,,
\end{equation}
where $\Lambda$ is the contraction with the  bivector field $(\omega_X^{1,1})^{-1}$, and $F_H = (F_{H_v})$ is the curvature corresponding to the metric $H=H_v$, $\forall v \in Q_0$.   Here the compositions on the l.h.s.  are defined as 
\begin{align}
 \phi_a \circ \phi^{\star H_a}_a : \mathcal E_{s(a)} \to M_a\otimes \mathcal R_{t(a)}\to \mathcal E_{s(a)}\,,\qquad \phi^{\star H_a}_a \circ \phi_a:\mathcal E_{t(a)}\to M_a^*\mathcal E_{s(a)}\to \mathcal E_{t(a)}\,.
\end{align}

A special case of the above vortex equation is when $Q$ is one  vertex $v$ with one loop $a$, and map $M_a\otimes \mathcal E_v\to \mathcal E_v$ gives a map from $M:=M_a$ to \emph{commuting} endomorphisms of $\mathcal E:=\mathcal E_v$.
 Such an object can be interpreted as a coherent sheaf on the total space of the dual bundle $M^*$ with $\dim_\C X$-dimensional support which is proper and finite over $X$, and such that the direct image to $X$ is a vector bundle.
  In the case $M=T_X$ this is equivalent to the  Hitchin equation. When $\dim_\C X=2$ and $M=\wedge^2 T^*_X$ we get Vafa-Witten equation, and when $n=\dim_\C X>2$ and $M=\wedge^n T^*_X$ we get a generalization of Vafa-Witten equations considered by one of us (G.B.) in an unpublished manuscript.
  In all these examples the total space of $M^*$ is a non-compact Calabi-Yau space in the algebro-geometric sense, i.e. it is endowed  with a non-vanishing holomorphic volume form.

\begin{remark}
For any quiver $Q$ and a collection of bundles $M_a$ labeled by the arrows of $Q$ one can construct a new quiver $Q'$ with the same set of vertices $Q_0'=Q_0$ and with exactly one edge $a'_{ij}$ for every ordered pair $(i,j)$ of vertices. The new bundles $M_{ij}'$ can be defined as the direct sums
\begin{align}
    M_{ij}':=\oplus_{a\in Q_1: s(a)=i,t(a)=j} M_a\,.
\end{align}
There is an obvious equivalence between the $M$-twisted $Q$-bundles and the $M'$-twisted $Q'$-bundles, and the corresponding harmonic metrics. Nevertheless, for  bookkeeping purposes, it is more convenient to work with the original description.
\end{remark}

\subsubsection{Nahm's equation} The algebra $C$ is $C^\infty(X)$ where $X$ is a $1$-dimensional manifold. The bimodule is supported on the diagonal and is $\Gamma(X,T^*_X\otimes_\R\C)\oplus C^\infty_\C(X)$, looks like the tensor product of $1$-forms on $X$ and the quiver with one vertex and one loop. The equation for harmonic representatives is exactly Nahm equation for the group $U(k)$: $\dot{A}_i=\epsilon_{ijk}[A_j,A_k]$ where  $ A_i=-A_i^\dagger\in Mat(k\times k,\C)$ are functions of time.

\subsection{Noncommutative instantons}

Ignoring the problem related to the \emph{noncompactness} of the noncommutative space $\R^{2n}_\theta$, the corresponding framework is the following. The algebra $\A$ is certain $C^\infty$-version of the algebra generated by generators $z_1,\dots,z_n$ and their Hermitian conjugates $z_1^*,\dots,z_n^*$
 satisfying relations\footnote{One can further generalize these relations and get \emph{holomorphic noncommutative spaces}, via replacing \eqref{commhol} by $[z_i,z_j]=c_{ij}$ and $[z_i^*,z_j^*]=-\cc{c}_{ij}$ where $(c_{ij})_{1\le i,j\le n}$ is any  skew-symmetric complex $n\times n$ matrix.}
 \begin{align}
  \label{commhol}   [z_i,z_j]=0,\,\,[z_i^*,z_j^*]=0\,,\\
     [z_i^*,z_j]=\hbar \delta_{ij}\,.
 \end{align}
 
 The algebra $\A$ is endowed with commuting derivations $\partial_1,\dots,\partial_n$ and $\cc{\partial}_1,\dots,\cc{\partial}_n$ given by
 
 \begin{gather}
     \partial_i(z_j)=\cc{\partial}_i(z_j^*)=\delta_{ij}\,,\\
     \partial_i(z_j^*)=\cc{\partial}_i(z_j)=0\,.
 \end{gather}

A noncommutative HYM instanton is a finitely-generated projective $\A$-module $E$ endowed with a $\A$-valued Hermitian form (see \eqref{notionhermitian})
\begin{align}
    \mathsf H:E \otimes_\C \overline{E}\to \A,\quad \mathsf H(f \phi_1,\cc{g\cdot \phi_2})=f\cdot \mathsf H(\phi_1,\cc{\phi_2})\cdot g^*
\end{align}
endowed with $\C$-linear endomorphisms $\nabla_1,\dots,\nabla_n$ and $\cc{\nabla}_1,\dots,\cc{\nabla}_n$
satisfying relations
\begin{gather}
[\nabla_i,\nabla_j]=[\cc{\nabla}_i,\cc{\nabla}_j]=0 \,,\\
[\nabla_i,z_j]=[\cc{\nabla}_i,\cc{z}_j]=\delta_{ij}\label{rel2}\,,\\
[\nabla_i,\cc{z}_j]=[\cc{\nabla}_i,z_j]=0\,,\\
\sum_{i=1}^n[\cc{\nabla}_i,\nabla_i]=0\label{rel4}\,,
\end{gather}
and
\begin{align}\label{leib}
     \mathsf H(\cc{\nabla}_i(\phi_1),\cc{\phi_2})+\mathsf H(\phi_1,\cc{{\nabla}_i(\phi_2)})=\cc{\partial}_i(\mathsf H(\phi_1,\cc{\phi_2}))\,.
    \end{align}
    
    \ 
    
\subsection{From noncommutative HYM to infinite-dimensional King's equation}\label{HYMtoKing}
The algebra $\A$ has a positive functional (state) $\int_\rho:\A\to \C$ (depending on arbitrary constant $\rho>0$) satisfying
\begin{align}
    \int_\rho  a  a^* \ge 0\quad\forall a\in \A
\end{align}
and given by 
\begin{align}\label{explicit}
    \int_\rho \prod_i z_i^{k_i} \prod_i (z_i^*)^{l_i}=
    \prod_{i=1}^n \delta_{k_i,l_i} k_i!\rho^{k_i}\,.
\end{align}

One can check using \eqref{explicit} that one has $\forall a\in \A,\,\forall i \in \{1,\dots,n\}$:
\begin{align}\label{byparts}
    \int_\rho \cc{\partial}_i(a)&=\frac{1}{\rho +\hbar}\int_\rho  a\cdot z_i \,,\\
    \int_{\rho} z_i \cdot a&=\frac{\rho}{\rho+\hbar}\int_\rho a\cdot z_i \label{exch}\,.
\end{align}
Let us introduce a non-negative \emph{$\C$-valued} pre-Hermitian pairing  on $E$ by 
\begin{align}
    \langle\phi_1,\phi_2\rangle:= \int_\rho\mathsf H(\phi_1,\cc{\phi_2})\,.
\end{align}
We conclude from \eqref{leib} and \eqref{byparts} that
\begin{multline}\label{dbarconj}
    \langle \cc{\nabla}_i \phi_1,\phi_2\rangle=\int_\rho \mathsf  H(\cc{\nabla}_i(\phi_1),\cc{\phi_2})=
\int_\rho \cc{\partial}_i(\mathsf H(\phi_1,\cc{\phi_2}))- \int_\rho \mathsf H(\phi_1,\cc{{\nabla}_i(\phi_2)})=\\
= \frac{1}{\rho+\hbar}\int_{\rho}\mathsf H(\phi_1,\cc{\phi_2})\cdot z_i-\langle \phi_1,\nabla_i(\phi_2)\rangle=\langle \phi_1,\frac{1}{\rho+\hbar}z_i^*\cdot \phi_2-\nabla_i(\phi_2)\rangle\,.
\end{multline}
Also, it follows from \eqref{exch} that 
\begin{align}\label{multconj}
    \langle z_i\cdot \phi_1,\phi_2\rangle= \int_\rho \mathsf H(z_i\cdot \phi_1,\phi_2)=\int_\rho z_i\cdot \mathsf H(\phi_1,\phi_2)=\frac{\rho}{\rho+\hbar}\int_\rho \mathsf H(\phi_1,\phi_2)\cdot z_i=\langle \phi_1,\frac{\rho}{\rho+\hbar}z^*_i\cdot \phi_2\rangle\,.
\end{align}
Let us introduce operators  in the Hilbert space $\mathcal H$  which is the completion of $E$ with respect to $\langle \cdot,\cdot\rangle$:
\begin{align}
    Z_i=z_i-\rho \cc{\nabla}_i\,.
\end{align}
The equations \eqref{dbarconj} and \eqref{multconj} imply that 
\begin{align}
    Z_i^\dagger=\rho \nabla_i\,.
\end{align}
Finally, using \eqref{rel2} and \eqref{rel4} we conclude that 
\begin{align}\label{infKing}
    \sum_{i=1}^n[Z_i^\dagger, Z_i]=\rho \cdot n\cdot \id_{\mathcal H}\,.
\end{align}
Consider the subspace $\mathcal H_0\subset \mathcal H$ which the common kernel of operators $\cc{\nabla}_i, \,\,i=1,\dots,n$. This subspace is preserved by the operators $z_i$, hence it is preserved by the operators $Z_i$.
We claim (the argument in not totally rigorous) that $\mathcal H_0\subset \mathcal H$  is also preserved by the  adjoint operators $Z_i^\dagger$. Indeed, it is the case when $E$ is the trivial bundle of rank one (in this case $\mathcal H_0$ is a completion of $\C[z_1,\dots,z_n]$). In general, let us consider the orthogonal decomposition 
\begin{align}
    \mathcal  H=\mathcal H_0\oplus \mathcal H_1,\quad \mathcal H_1:=\mathcal H_0^\perp\,.
\end{align}
In this splitting we have for any $i=1,\dots ,n$:
\begin{align}
    Z_i=\begin{pmatrix}Z_i^{00} & Z_i^{01}\\ 0 & Z_i^{11}\end{pmatrix},\quad Z_i^\dagger=\begin{pmatrix}(Z_i^{00})^\dagger & 0 \\ (Z_i^{01})^\dagger & (Z_i^{11})^\dagger\end{pmatrix}\,.
\end{align}
We conclude that
\begin{align}\label{Zidentity}
    \sum_i [(Z_i^{00})^\dagger, Z_i^{00}]+\sum_i (Z_i^{01})^\dagger Z_i^{(01)}=\rho\cdot n\cdot \id_{\mathcal H_0}\,.
\end{align}
For each $i$ the operator $[(Z_i^{00})^\dagger, Z_i^{00}]-\rho\cdot \text{Id}_{\mathcal H_0}$ is of trace class, \emph{hence} its trace is equal to zero (reasoning: the trace does not change by small deformations). Together with \eqref{Zidentity} this vanishing of traces implies that
\begin{align}
\sum_i \text{Trace}((Z_i^{01})^\dagger Z_i^{(01)})=0\,,
\end{align}
and therefore \emph{all} operators $Z_i^{(01)}$ vanish.  Hence, the equation \eqref{infKing} holds on $\mathcal H_0$ as well. This concludes the argument.

\section{Acknowledgements}
The first author acknowledges, the financial support from  Fondation CFM pour la Recherche, Institut des Hautes \'{E}tudes Scientifiques for their hospitality. He thanks his first mathematics teacher in high school, Shri Susanta Kumar Brahma, he also thanks  Professor Ranendra Narayan Biswas, Professor Mandar Mitra, Professor Athanase Papadopoulos, Professor Yuri Manin, Professor Igor Volovich, Professor Francois Laudenbach, Professor Valentin Poenaru, and Professor Pierre Cartier for useful discussions and help of various kinds.

The second author thanks Nikita Nekrasov for the introducing him to the beautiful equation [\ref{infKing}] many years ago.

Both authors thank Frederic Paulin for the remarks he made and his interest to the paper.

\end{document}